\documentclass[aps,prd,nofootinbib,showkeys,preprint,floatfix]{revtex4}

\usepackage{amsmath}
\usepackage{amssymb}
\usepackage{graphicx}
\usepackage{rotating}
\usepackage{color} 
\usepackage{multirow} 
 \usepackage[T1]{fontenc}

\newcommand{\AddrAHEP}{
  {\it AHEP Group, Instituto de F\'{\i}sica Corpuscular --
    C.S.I.C./Universitat de Val{\`e}ncia \\
    Edificio de Institutos de Paterna, Apartado 22085,
  E--46071 Val{\`e}ncia, Spain}}

\newcommand{\AddrLisb}{%
 Departamento de F\'\i sica and CFTP, Instituto Superior T\'ecnico\\
          Av. Rovisco Pais 1, 1049-001 Lisboa, Portugal }

\newcommand{\AddrWur}{%
Institut f\"ur Theoretische Physik und Astronomie, 
Universit\"at W\"urzburg\\
Am Hubland, 
97074 Wuerzburg}

\def\gsim{\raise0.3ex\hbox{$\;>$\kern-0.75em\raise-1.1ex\hbox{$\sim\;$}}}
\def\lsim{\raise0.3ex\hbox{$\;<$\kern-0.75em\raise-1.1ex\hbox{$\sim\;$}}}

\begin{document}

\preprint{CFTP/09-029}  
\preprint{IFIC/09-35}

\title{Dark matter in minimal supergravity with type-II seesaw}

\author{J.~N.~Esteves}\email{joaomest@cftp.ist.utl.pt}\affiliation{\AddrLisb}
\author{M.~Hirsch} \email{mahirsch@ific.uv.es}\affiliation{\AddrAHEP}
\author{S.~Kaneko} \email{satoru@cftp.ist.utl.pt}\affiliation{\AddrLisb}
\author{W. Porod} \email{porod@physik.uni-wuerzburg.de}\affiliation{\AddrWur}
\author{J.~C.~Romao}\email{jorge.romao@ist.utl.pt}\affiliation{\AddrLisb}



\begin{abstract}
We calculate the relic density of the lightest neutralino in a 
supersymmetric seesaw type-II (``triplet seesaw'') model with 
minimal supergravity boundary conditions at the GUT scale. The 
presence of a triplet below the GUT scale, required to explain 
measured neutrino data in this setup, leads to a characteristic 
deformation of the sparticle spectrum with respect to the pure 
mSugra expectations, affecting the calculated relic dark matter (DM) 
density. We discuss how the DM allowed regions in the ($m_0,M_{1/2}$) 
plane change as a function of the (type-II) seesaw scale. We 
also compare the constraints imposed on the models parameter 
space form upper limits on lepton flavour violating (LFV) decays 
to those imposed by DM. Finally, we briefly comment on uncertainties 
in the calculation of the relic neutralino density due to uncertainties 
in the measured top and bottom masses. 

\end{abstract}

\maketitle

\section{Introduction}
\label{sec:int}

Standard cosmology requires the existence of a non-baryonic dark 
matter (DM) contribution to the total energy budget of the universe 
\cite{Jungman:1995df,Bertone:2004pz}. In the past few years estimates 
of the DM abundance have become increasingly precise. Indeed, the 
Particle Data Group now quotes at 1 $\sigma$ c.l. \cite{PDG2008}
\begin{equation}\label{OmCDM}
\Omega_{DM}h^2 = 0.105\pm 0.008.
\end{equation}
Since the data from the WMAP satellite \cite{Spergel:2003cb,Komatsu:2008hk} 
and large scale structure formation \cite{Tegmark:2006az} is best fitted 
if the DM is cold, weakly interacting mass particles (WIMP) are 
currently the preferred explanation. While there is certainly no shortage 
of WIMP candidates (lists can be found in many reviews, see for example 
\cite{Jungman:1995df,Bertone:2004pz,Bergstrom:2009ib,Baer:2009bu}), the 
literature is completely dominated by studies of the lightest neutralino.

Neutrino oscillation experiments have shown that neutrinos have 
non-zero mass and mixing angles \cite{Fukuda:1998mi,Ahmad:2002jz,%
Eguchi:2002dm,Collaboration:2007zza,KamLAND2007} and the most recent 
global fits to all data \cite{Maltoni:2004ei} confirm again that 
the mixing angles are surprisingly close to the so-called tri-bimaximal 
mixing (TBM) values \cite{Harrison:2002er}. In the minimal 
supersymmetric extension of the standard model (MSSM) with conserved 
R-parity neutrino masses are zero for the same reasons as in the SM. 
However, it was shown long ago that if neutrinos are Majorana 
particles, their mass is described by a unique dimension-5 operator 
\cite{Weinberg:1979sa}
\begin{equation}\label{eq:dim5}
m_{\nu} = \frac{f}{\Lambda} (H L) (H L) .
\end{equation}
All (Majorana) neutrino mass models reduce to this operator at low 
energies. If $f$ is a coefficient ${\cal O}(1)$, current neutrino 
data indicates $\Lambda \lsim {\cal O}(10^{15})$ GeV. This is the 
essence of the ``seesaw'' mechanism. There are three different tree-level 
realizations of the seesaw, classified as type-I, type-II and type-III 
in \cite{Ma:1998dn}. Type-I is the well-known case of the exchange of 
a heavy fermionic singlet \cite{Minkowski:1977sc,seesaw,MohSen}. 
Type-II corresponds to the exchange of a scalar triplet 
\cite{Schechter:1980gr,Cheng:1980qt}. One could also add one (or more) 
fermionic triplets to the field content of the SM \cite{Foot:1988aq}. 
This is called seesaw type-III in  \cite{Ma:1998dn}.

Neutrino experiments at low energies measure only $f_{\alpha\beta}/\Lambda$, 
thus observables outside the neutrino sector will ultimately be needed 
to learn about the origin of eq. (\ref{eq:dim5}). Augmenting the SM 
with a high-scale seesaw mechanism does not lead to any conceivable 
phenomenology apart from neutrino masses, but if weak scale supersymmetry 
exists {\em indirect} probes into the high energy world might be 
possible. Two kind of measurements containing such indirect information 
exist in principle, lepton flavour violating (LFV) observables and sparticle 
masses. 

Assuming complete flavour blindness in the soft supersymmetry breaking 
parameters at some large scale, the neutrino Yukawa matrices will, in 
general, lead to non-zero flavour violating entries in the slepton 
mass matrices, if the seesaw scale is lower than the scale at which 
SUSY is broken. This was first pointed out in \cite{Borzumati:1986qx}. 
The resulting LFV processes have been studied in many publications, 
for low-energy observables such as $\mu \to e \gamma$ and $\mu-e$ 
conversion in seesaw type-I see for example
\cite{Hisano:1995nq,Hisano:1995cp,Ellis:2002fe,Deppisch:2002vz,%
Arganda:2005ji,Antusch:2006vw,Arganda:2007jw}, 
for seesaw type-II \cite{Rossi:2002zb,Hirsch:2008gh}. 
LFV collider observables have also been studied in a number of papers, 
see for example \cite{Hirsch:2008gh,Hisano:1998wn,Krasnikov:1995qq,%
ArkaniHamed:1996au,Nomura:2000zb,Porod:2002zy,Deppisch:2003wt,%
delAguila:2008iz,Deppisch:2005rv,Bartl:2005yy,Hirsch:2008dy,%
Carquin:2008gv,Esteves:2009vg}.

Mass measurements in the sparticle sector will not only be necessary 
to learn about the mechanism of SUSY breaking in general, but might 
also reveal indications about the scale of the seesaw mechanism. However, 
very precise knowledge of masses will be necessary before one can learn 
about the high scale parameters \cite{Blair:2002pg,Freitas:2005et}. 
Especially interesting in this context is the observation that from 
the different soft scalar and gaugino masses one can define certain 
combinations (``invariants'') which are nearly constant over large 
parts of mSugra space. Adding a seesaw mechanism of type-II or type-III 
these invariants change in a characteristic way as a function of the 
seesaw scale and are thus especially suited to extract information 
about the high energy parameters \cite{Buckley:2006nv}. Note, however, 
that the ``invariants'' are constants in mSugra space only in leading 
order and that quantitatively important 2-loop corrections 
exist \cite{Hirsch:2008gh}.

In this paper we study neutralino dark matter 
\cite{Ellis:1983ew,Griest:1990kh,Drees:1992am} within a supersymmetric 
type-II seesaw model with mSugra boundary conditions. For definiteness, 
the model we consider consists of the MSSM particle spectrum to which 
we add a single pair of ${\bf 15}$- and $\bf\overline{15}$-plets. This 
is the simplest supersymmetric type-II setup, which allows one to maintain 
gauge coupling unification \cite{Rossi:2002zb} and explain measured 
neutrino oscillation data.

In mSugra -  assuming a standard thermal history of the early universe 
\footnote{In models with non-standard thermal history the relation 
between sparticle masses and relic density can be lost completely 
\cite{Gelmini:2006pw}.} - only four very specific regions in parameter 
space can correctly explain the most recent WMAP data \cite{Komatsu:2008hk}. 
These are (i) the bulk region; (ii) the co-annihilation line; (iii) the 
``focus point'' line and (iv) the ``higgs funnel'' region. In the bulk 
region there are no specific relations among the sparticle masses. However, 
all sparticles are rather light in this region, so it is already very 
constrained from the view point of low-energy data \cite{Allanach:2004xn}. 
In the co-annihilation line the lightest scalar tau is nearly degenerate 
with the lightest neutralino, thus reducing the neutralino relic density 
with respect to naive expectations \cite{Griest:1990kh,Baer:2002fv}. 
In the ``focus point'' line \cite{Feng:1999zg,Baer:2002fv} 
$\Omega_{\chi^0_1}h^2$ is small enough to explain $\Omega_{DM}h^2$ 
due to a rather small value of $\mu$ leading to an enhanced higgsino 
component in the lightest neutralino and thus an enhanced coupling 
to the $Z^0$ boson. Lastly, at large $\tan\beta$ an s-channel resonance 
pair annihilation of neutralinos through the CP-odd higgs boson can become 
important. This is called the ``higgs funnel'' region \cite{Drees:1992am}.

The addition of the ${\bf 15}$ and $\bf\overline{15}$ pair at the high 
scale does not, in general, lead to the appearance of new allowed regions. 
However, the deformed sparticle spectrum with respect to mSugra expectations 
leads to characteristic changes in the allowed regions as a function 
of the unknown seesaw scale. We discuss these changes in detail and 
compare the results to other indirect constraints, namely, the observed 
neutrino masses and upper limits on LFV processes. We concentrate on 
the seesaw type-II scheme, since for mSugra + seesaw type-I the changes 
in the DM allowed regions with respect to pure mSugra are, in general, 
expected to be tiny. \footnote{We have confirmed this general expectation 
with some sample calculations. However, an exceptional case has been 
presented recently in \cite{Kadota:2009vq}, see the more detailed discussion 
in section (\ref{sec:results}).} 

The rest of this paper is organized as follows. In the next section 
we briefly summarize the main ingredients of the model and give a 
short discussion of mSugra and the expected changes in sparticle 
masses in our setup with respect to mSugra. In section (\ref{sec:results}) 
we present our numerical results. This is the main section of the 
current paper, where we discuss in detail how the introduction of a 
${\bf 15}$ changes the predicted DM abundance as a function of the 
seesaw scale. We also confront the DM allowed regions with constraints 
from non-observation of LFV processes and briefly comment on DM in 
mSugra with a seesaw type-I. We then close with a short summarizing 
discussion in section (\ref{sec:cncl}).

\section{Setup: mSugra and $SU(5)$ motivated type-II seesaw}
\label{sec:model}

In this section we summarize the main features of the model 
we will use in the numerical calculation. We will always refer 
to minimal Supergravity (mSugra) as the ``standard'' against 
which we compare all our results. The model consists in 
extending the MSSM particle spectrum by a pair of ${\bf 15}$ 
and $\bf\overline{15}$. It is the minimal supersymmetric seesaw 
type-II model which maintains gauge coupling unification 
\cite{Rossi:2002zb}.

mSugra is specified by 4 continuous and one discrete parameter 
\cite{Nilles:1983ge}. These are usually chosen to be $m_0$, the 
common scalar mass, $M_{1/2}$, the gaugino mass parameter, $A_0$, 
the common trilinear parameter, $\tan\beta=\frac{v_2}{v_1}$ and 
the sign of $\mu$. $m_0$, $M_{1/2}$ and $A_0$ are defined at the 
GUT scale, the RGEs are known at the 2-loop level \cite{Martin:1993zk}.

Under $SU(3)\times SU_L(2) \times U(1)_Y$ the ${\bf 15}$ decomposes as 
\begin{eqnarray}\label{eq:15}
{\bf 15} & = &  S + T + Z \\ \nonumber
S & \sim  & (6,1,-\frac{2}{3}), \hskip10mm
T \sim (1,3,1), \hskip10mm
Z \sim (3,2,\frac{1}{6}).
\end{eqnarray}
The $SU(5)$ invariant superpotential reads as 
\begin{eqnarray}\label{eq:pot15}
W & = & \frac{1}{\sqrt{2}}{\bf Y}_{15} {\bar 5} \cdot 15 \cdot {\bar 5} 
   + \frac{1}{\sqrt{2}}\lambda_1 {\bar 5}_H \cdot 15 \cdot {\bar 5}_H 
+ \frac{1}{\sqrt{2}}\lambda_2 5_H \cdot \overline{15} \cdot 5_H 
+ {\bf Y}_5 10 \cdot {\bar 5} \cdot {\bar 5}_H \\ \nonumber
 & + & {\bf Y}_{10} 10 \cdot 10 \cdot 5_H + M_{15} 15 \cdot \overline{15} 
+ M_5 {\bar 5}_H \cdot 5_H
\end{eqnarray}
Here, ${\bar 5}=(d^c,L)$, $10=(u^c,e^c,Q)$, ${5}_H =(t,H_2)$ and 
${\bar 5}_H=({\bar t},H_1)$. Below the GUT scale in the $SU(5)$-broken 
phase the potential contains the terms 
\begin{eqnarray}\label{eq:broken}
 &  & \frac{1}{\sqrt{2}}(Y_T L T_1 L +  Y_S d^c S d^c) 
+ Y_Z d^c Z L + Y_d d^c Q H_1 + Y_u u^c Q H_2  
+ Y_e e^c L H_1 \\ \nonumber
& + & \frac{1}{\sqrt{2}}(\lambda_1 H_1 T_1 H_1 +\lambda_2 H_2 T_2 H_2) 
+ M_T T_1 T_2 + M_Z Z_1 Z_2 + M_S S_1 S_2 + \mu H_1 H_2 
\end{eqnarray}
$Y_d$, $Y_u$ and $Y_e$ 
generate quark and charged lepton masses in the usual manner. In addition 
there are the matrices $Y_T$, $Y_S$ and $Y_Z$. For the case of a complete 
${\bf 15}$, apart from calculable threshold corrections, $Y_T=Y_S=Y_Z$ 
and $M_T$, $M_S$ and $M_Z$ are determined from $M_{15}$ by the RGEs.
As long as $M_Z \sim M_S \sim M_T \sim M_{15}$ gauge coupling unification 
will be maintained. The equality need not be exact for successful unification. 

The triplet $T_1$ has the correct quantum numbers to generate neutrino 
masses via the first term in eq. (\ref{eq:broken}). Integrating out the 
heavy triplets at their mass scale a dimension-5 operator of the form 
eq. (\ref{eq:dim5}) is generated and after electro-weak 
symmetry breaking the resulting neutrino mass matrix can be written as 
\begin{eqnarray}\label{eq:ssII}
m_\nu=\frac{v_2^2}{2} \frac{\lambda_2}{M_T}Y_T.
\end{eqnarray}
Here $v_2$ is the vacuum expectation value of Higgs doublet $H_2$ and 
we use the convention $\langle H_i\rangle = \frac{v_i}{\sqrt{2}}$. 
$m_\nu$ can be diagonalized in the standard way with a unitary matrix 
$U$, containing in general 3 angles and 3 phases. Note that 
${\hat Y}_T = U^T \cdot Y_T \cdot U$ is diagonalized by {\em the same 
matrix as $m_{\nu}$}. This means that if all neutrino eigenvalues, angles 
and phases were known, $Y_T$ would be completely fixed up to an overall 
constant, which can be written as 
$\frac{M_T}{\lambda_2} \simeq 10^{15} {\rm GeV} \hskip2mm 
\Big(\frac{0.05 \hskip1mm {\rm eV}}{m_{\nu}}\Big)$. 
Thus, current neutrino data requires $M_T$ to be lower than the GUT scale 
by (at least) an order or magnitude.

The full set of RGEs for the ${\bf 15}$ + $\overline{\bf 15}$ can 
be found in \cite{Rossi:2002zb} and in the numerical calculation, 
presented in the next section, we solve the exact RGEs. However, 
for a qualitative understanding of the results, the following 
approximative solutions are quite helpful. 

For the gaugino masses one finds in leading order 
\begin{eqnarray}
M_i(m_{SUSY}) = \frac{\alpha_i(m_{SUSY})}{\alpha(M_G)} M_{1/2}.
\label{eq:gaugino}
\end{eqnarray}
Eq. (\ref{eq:gaugino}) implies that the ratio $M_2/M_1$, which is 
measured at low-energies, has the usual mSugra value, but the 
relationship to $M_{1/2}$ is changed.
Neglecting the Yukawa couplings ${\bf Y}_{15}$ (see below), for the 
soft mass parameters of the first two generations one gets 
\begin{eqnarray}\label{eq:scalar}
m_{\tilde f}^2  &=& M_0^2  +  
\sum_{i=1}^3  c^{\tilde f}_i \left(
\left(\frac{\alpha_i(M_T)}{\alpha(M_G)}\right)^2 f_i
 + f_i'
\right) M_{1/2}^2, \\ \nonumber
 f_i &=& \frac{1}{b_i} \left(
1 - {\left[1 + \frac{\alpha_i(M_T)}{4 \pi} b_i \log
\frac{M^2_T}{m_Z^2}\right]^{-2}} \right), \\
   f_i' &=& \frac{1}{b_i+\Delta b_i} \left(
1 - {\left[1 + \frac{\alpha(M_G)}{4 \pi} (b_i + \Delta b_i) \log
\frac{M_G^2}{M_T^2}\right]^{-2}} \right).
\end{eqnarray}
The various coefficients $c^{\tilde f}_i$ can be found 
in \cite{Hirsch:2008gh}. The gauge couplings are given as 
\begin{eqnarray}
\alpha_1(m_Z) &=& \frac{5 \alpha_{em}(m_Z)}{3 \cos^2\theta_W}, \hspace{1cm}
\alpha_2(m_Z) = \frac{\alpha_{em}(m_Z)}{\sin^2\theta_W}, \\  \nonumber 
\alpha_i(m_{SUSY}) &=& \frac{\alpha_i(m_Z)}{1- \frac{\alpha_i(m_Z)}
               {4 \pi} b_i^{SM} \log{\frac{m_{SUSY}^2}{m_Z^2}}}, \\ \nonumber
\alpha_i(M_T) &=& \frac{\alpha_i(m_{SUSY})}{1- \frac{\alpha_i(m_{SUSY})}
               {4 \pi} b_i \log{\frac{M_T^2}{m_{SUSY}^2}}}, \\ \nonumber
\alpha_i(M_G) &=& \frac{\alpha_i(M_T)}{1- \frac{\alpha_i(M_T)}
               {4 \pi} (b_i+\Delta b_i) \log{\frac{M_G^2}{M_T^2}}}.
\end{eqnarray}
with $b^{SM}_i$ and $b^{MSSM}_i$ being the usual standard model and MSSM 
coefficients. $\Delta b_i=7$ for all $i$ in case of a complete 15-plet.

\begin{figure}[!htb]
  \centering
  \begin{tabular}{cc}
\includegraphics[width=0.48\textwidth]{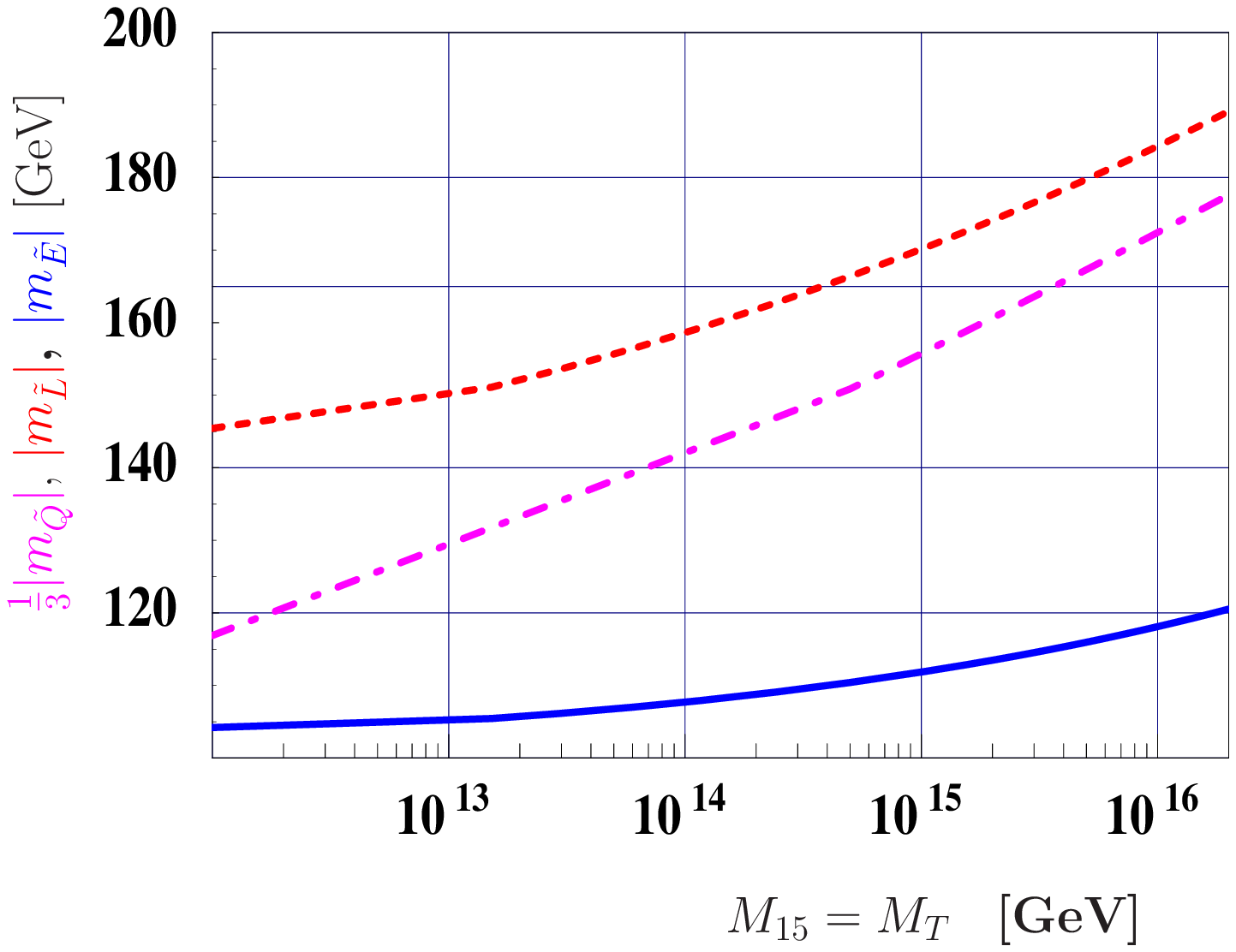}
\includegraphics[width=0.48\textwidth]{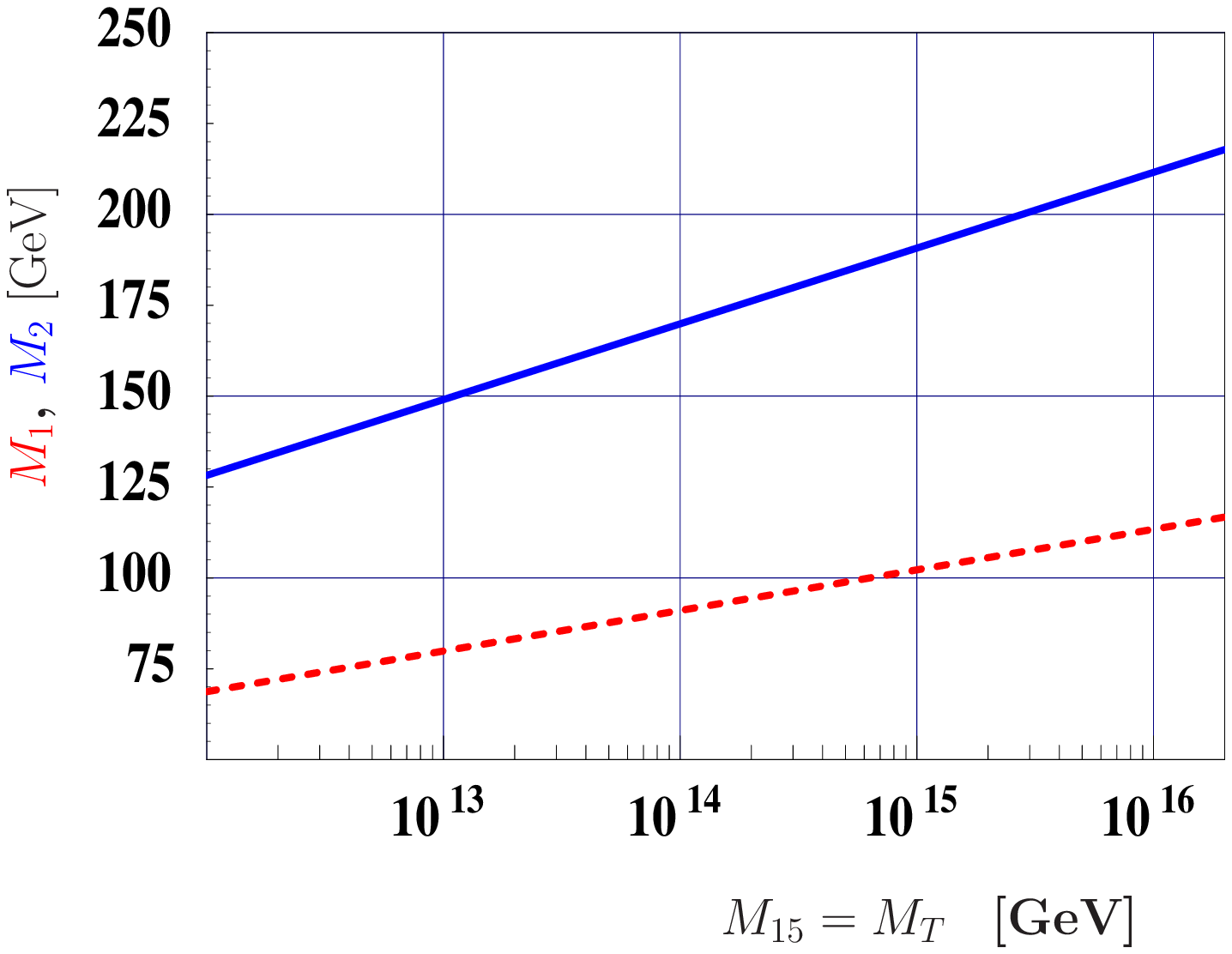}
\end{tabular}
\vspace{-5mm}
\caption{Analytically calculated running of scalar (to the left) 
and gaugino mass parameters (to the right), leading order only. 
The mass parameters are calculated 
as a function of $M_{\bf 15}$ for the mSugra parameters $m_0=70$ GeV 
and $M_{1/2} = 250$ GeV. For $M_{\bf 15}\simeq 2 \times 10^{16}$ GeV 
the mSugra values are recovered. Smaller $M_{\bf 15}$ lead to smaller 
soft masses in all cases. Note that the running is different for the 
different mass parameters with gaugino masses running faster than 
slepton mass parameters.}
  \label{fig:ana_para}
\end{figure}

We can estimate the soft mass parameters given the above formulas 
for a given choice of $m_0,M_{1/2}$ and $M_{\bf 15}=M_T$. We show 
some arbitrarily chosen examples in fig. (\ref{fig:ana_para}). Note that 
the result shown is approximate, since we are  (a) using the leading log 
approximation and (b) two loop effects are numerically important, especially 
for $m_{Q}$, but not included. The figure serves to show that for 
any $M_{\bf 15} < M_{GUT}$ the resulting mass parameters are always 
smaller than the mSugra expectations for the same choice of initial 
parameters ($m_0,M_{1/2}$). While the exact values depend on 
($m_0,M_{1/2}$) and on the other mSugra parameters, this feature is 
quite generally true in all of the ($m_0,M_{1/2}$) plane. Note, that the 
running is different for the different scalar mass parameters, but the 
ratio of the gaugino mass parameters $M_1/M_2$ always stays close to the 
mSugra expectation, $M_1 \simeq \frac{5}{3}\tan^2\theta_W M_2$. 

\section{Numerical results}
\label{sec:results}

In this section we discuss our numerical results. All the plots 
shown below are based on the program packages SPheno \cite{Porod:2003um} 
and micrOMEGAs \cite{Belanger:2004yn,Belanger:2006is}. We use SPheno 
V3 \cite{PorodWebpage}, including the RGEs for the  ${\bf 15}$ + 
$\overline{\bf 15}$ case \cite{Rossi:2002zb,Hirsch:2008gh} at the 
2-loop level for gauge couplings and gaugino masses and at one-loop 
level for the remaining MSSM parameters and the 15-plet parameters, 
for a discussion see \cite{Hirsch:2008gh}. For any given set of 
mSugra and 15-plet parameters SPheno calculates the supersymmetric 
particle spectrum at the electro-weak scale, which is then interfaced 
with micrOMEGAs2.2 \cite{micrOMEGAsWebpage} to calculate the relic 
density of the lightest neutralino, $\Omega_{\chi^0_1}h^2$. 

For the standard model parameters we use the PDG 2008 values \cite{PDG2008}, 
unless specified otherwise. As discussed below, especially important 
are the values (and errors) of the bottom and top quark masses, 
$m_b = 4.2 + 0.17 - 0.07$ GeV and $m_t = 171.2 \pm 2.1$ GeV. 
Note, the $m_t$  is understood to be the pole-mass and $m_b(m_b)$
is the $\overline{MS}$ mass. As 
the allowed range for $\Omega_{DM}h^2$ we always use the 3 $\sigma$ 
c.l. boundaries as given in \cite{PDG2008}, i.e. $\Omega_{DM}h^2 = 
[0.081,0.129]$. Note, however that the use of 1 $\sigma$ contours 
results in very similar plots, due to the small error bars.

\begin{figure}[!htb]
  \centering
  \begin{tabular}{cc}
\includegraphics[width=0.48\textwidth]{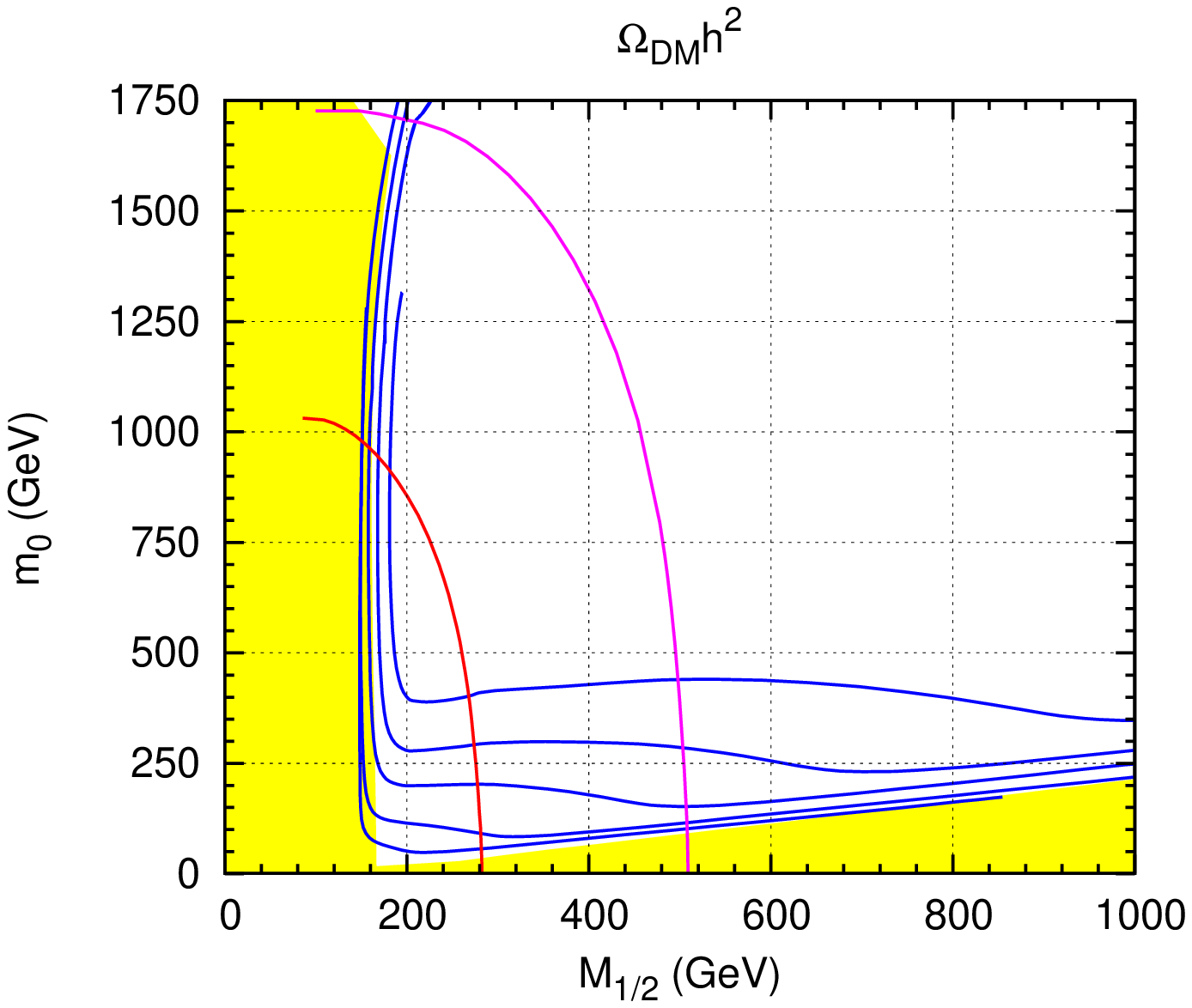}
\includegraphics[width=0.48\textwidth]{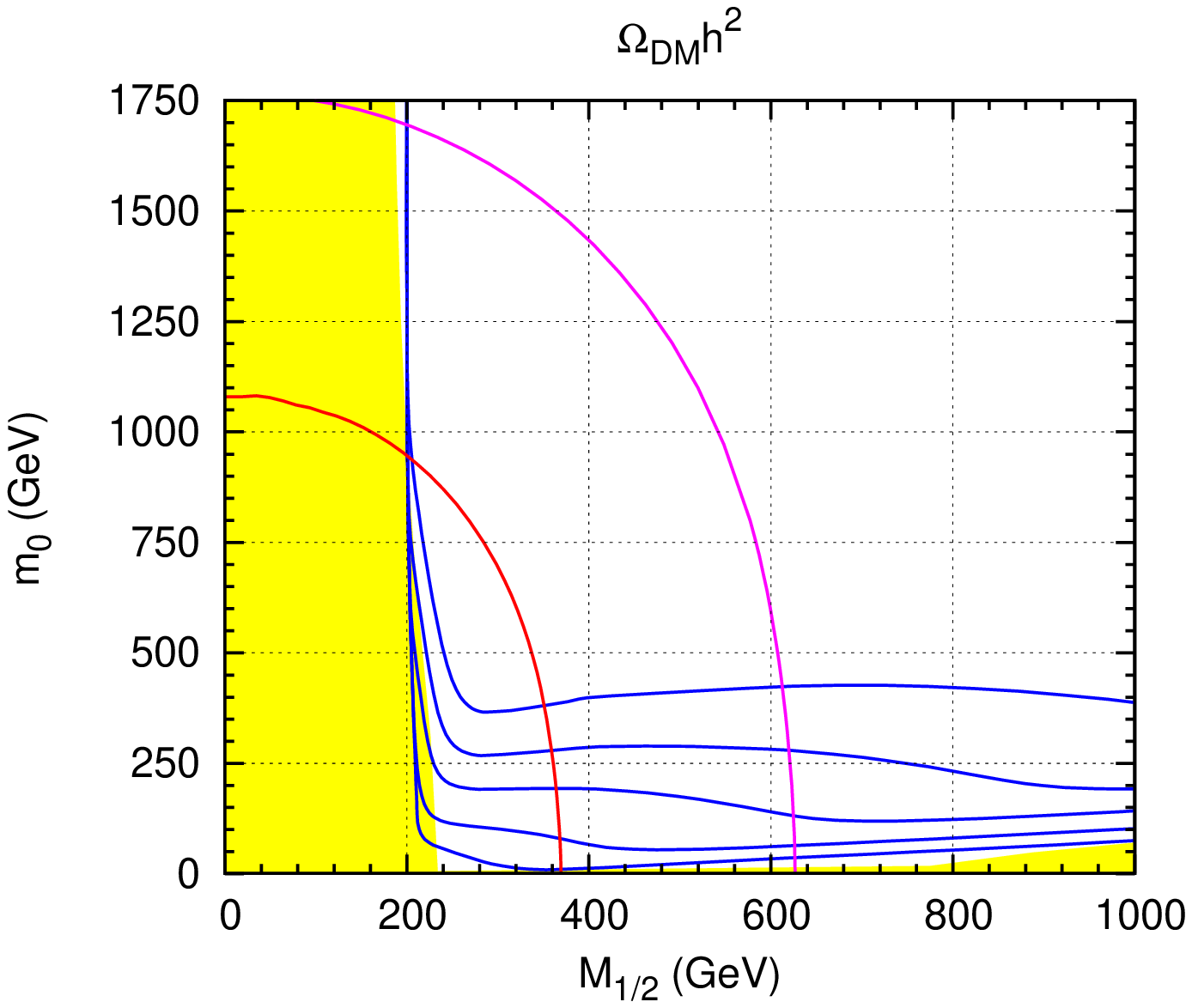} \\
 \includegraphics[width=0.48\textwidth]{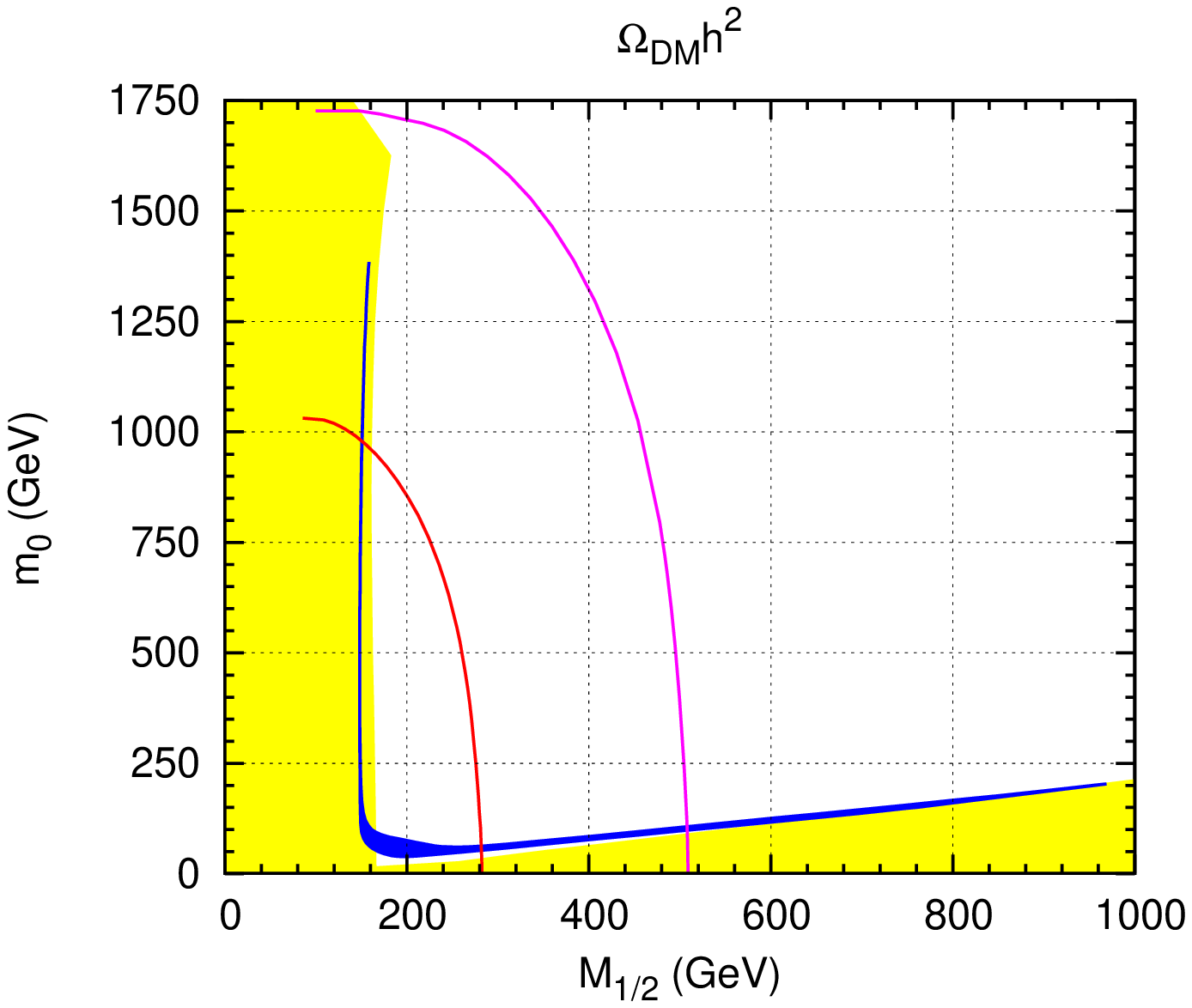}
 \includegraphics[width=0.48\textwidth]{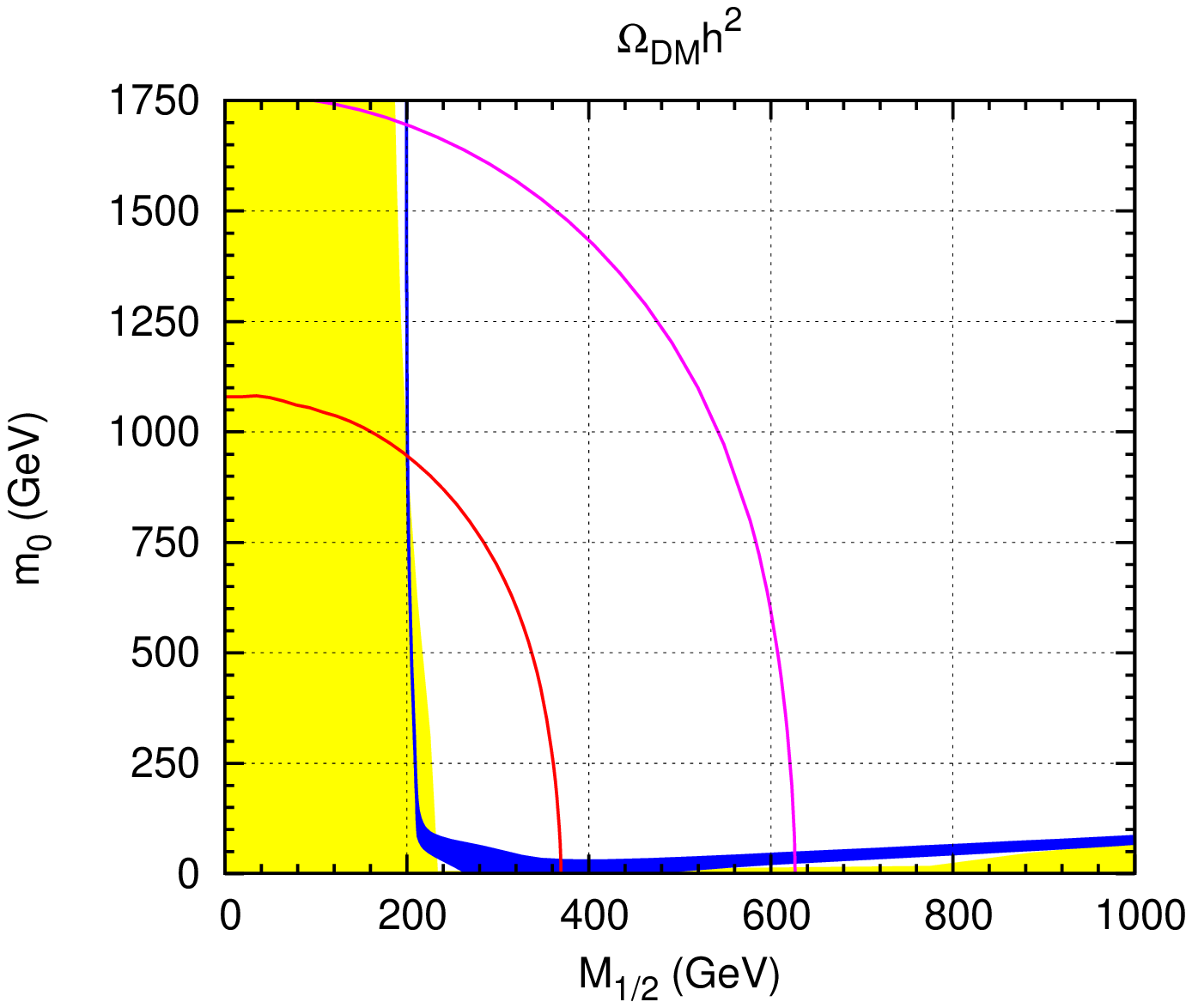}

\end{tabular}
\vspace{-5mm}
\caption{Top: Contours of equal dark matter density ($\Omega_{\chi^0_1} h^2$) 
in the ($m_0,M_{1/2}$) plane for the ``standard choice'' $\tan\beta=10$, 
$A_0=0$ and $\mu \ge 0$, for mSugra (left panel) and type-II seesaw
with $M_T=10^{14}$ GeV (right panel). The lines are constant 
$\Omega_{\chi^0_1} h^2$ with $\Omega_{\chi^0_1} h^2=0.1,0.2,0.5,1,2$. 
Bottom: Range of parameters allowed by the DM constraint at 3 $\sigma$ 
c.l. To the left: mSugra; to the right: $M_T=10^{14}$ GeV. For a 
discussion see text.}
\label{fig:standard}
\end{figure}

In the ``seesaw sector'' we have the parameters connected with the 15-plets, 
i.e. $M_{\bf 15}$, $Y_{\bf 15}$, $\lambda_1$ and $\lambda_2$. For the 
calculation of the dark matter abundance the most important parameter 
is $M_{\bf 15}$. It has turned out that the effects of $Y_{\bf 15}$, 
$\lambda_1$ and $\lambda_2$ on the relic abundance of neutralinos are 
very minor. Note, however, that as discussed in the previous section, 
atmospheric neutrino oscillation data can not be explained in our setup, 
if the triplet mass is larger than approximately $M_{\bf 15} = M_T = 10^{15}$ 
GeV. Also, the non-observation of lepton flavour violating (LFV) decays puts 
an upper bound on $M_{\bf 15}$. The latter, however, is strongly dependent 
on $\tan\beta$ and depends also on $m_0$ and $M_{1/2}$. We will first 
show results using different values of $M_T$ as free parameter, without 
paying attention to neutrino masses and LFV. We will discuss how our 
results change for correctly fitted neutrino masses and angles towards 
the end of this section, where we also discuss and compare LFV excluded 
regions with DM allowed ones.

We define our ``standard choice'' of mSugra parameters as $\tan\beta=10$, 
$A_0 =0$ and $\mu>0$ and use these values in all plots, unless specified 
otherwise. We then show our results in the plane of the remaining 
two free parameters, ($m_0,M_{1/2}$). Fig. (\ref{fig:standard}) shows 
in the top panel contours of equal dark matter density, 
$\Omega_{\chi^0_1} h^2$. The lines are constant $\Omega_{\chi^0_1} h^2$ 
with $\Omega_{\chi^0_1} h^2=0.1,0.2,0.5,1,2$. In the bottom panel we 
show the range of parameters allowed by the DM constraint at 3 $\sigma$ 
c.l. In both cases, to the left a pure mSugra calculation, whereas 
the plot to the right shows mSugra + 15-plet with $M_T=10^{14}$ GeV. 
In each plot the yellow regions are eluded either by the lighter 
scalar tau being the LSP (to the bottom right) or  by the LEP limit 
on the mass of the lighter chargino (to the left), 
$m_{\chi^+_1} \ge 105$ GeV. In addition, we show two lines of constant 
lightest Higgs boson mass, $m_{h^0} = 110$ GeV (dotted) and $m_{h^0}=114.4$ 
GeV (dashed), as calculated by SPheno, see the discussion below.

The plots show three of the different allowed regions discussed in 
the introduction. To the right the co-annihilation region, here 
the lightest neutralino and the lighter scalar tau are nearly degenerate 
in mass. The line going nearly vertically upwards at constant $M_{1/2}$ 
is the ``focus point'' line. The small region connecting the two lines 
are the remains of the bulk region, which has shrunk considerably due 
to the reduced error bars on $\Omega_{DM}h^2$ after the most recent 
WMAP data \cite{Komatsu:2008hk}. The focus point line is excluded 
by the LEP constraint on the lighter chargino mass at low and moderate 
values of $m_0$. It becomes allowed only at values of $m_0$ larger 
than (very roughly) 1-1.5 TeV. However, note that the exact value of $m_0$ 
at which the focus point line becomes allowed is extremely sensitive 
to errors in $m_{\chi^+_1}$, both from the experimental bound and the 
error in the theoretical calculation. 

\begin{figure}[!htb]
  \centering
  \begin{tabular}{cc}
\includegraphics[width=0.48\textwidth]{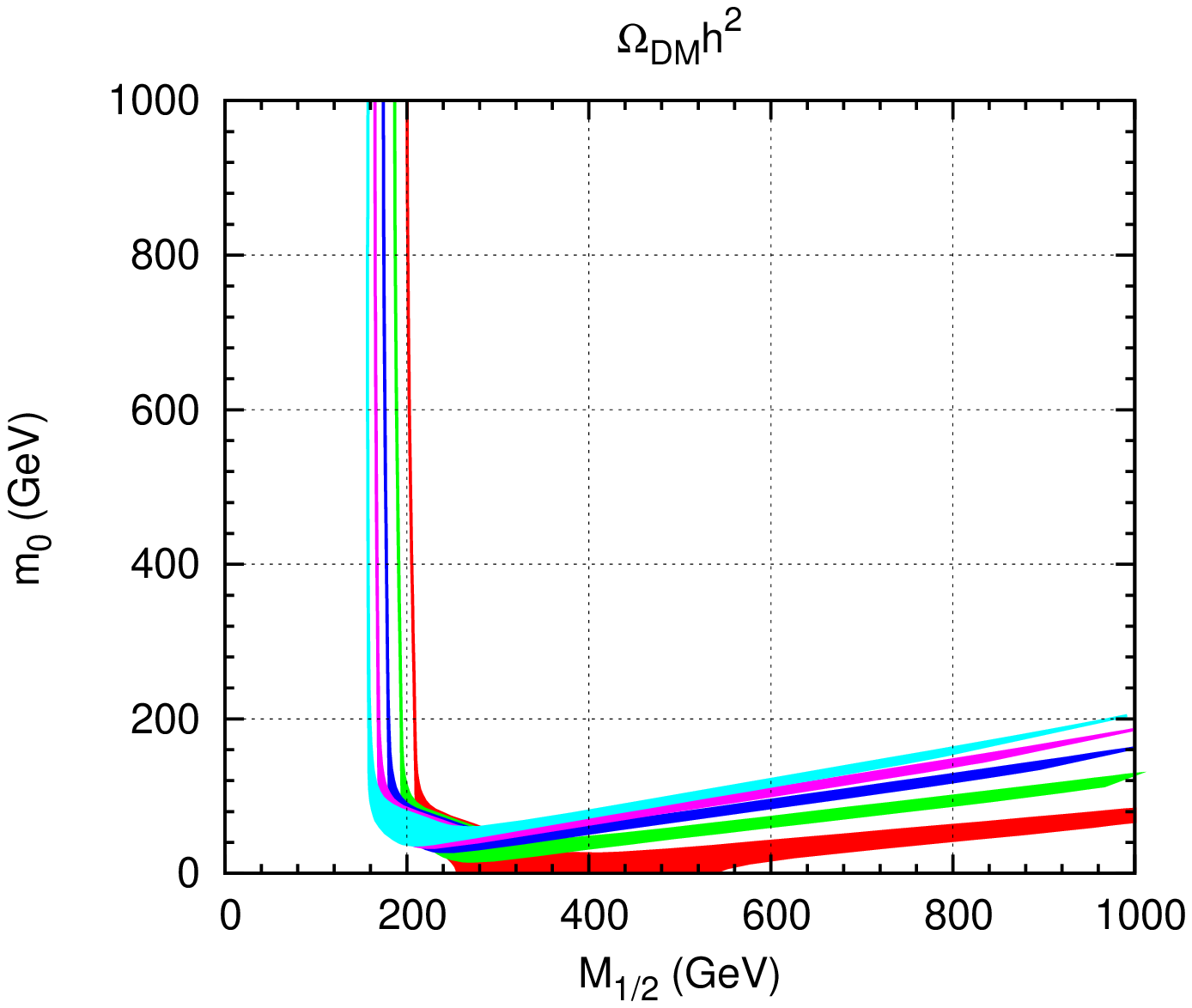}
\includegraphics[width=0.48\textwidth]{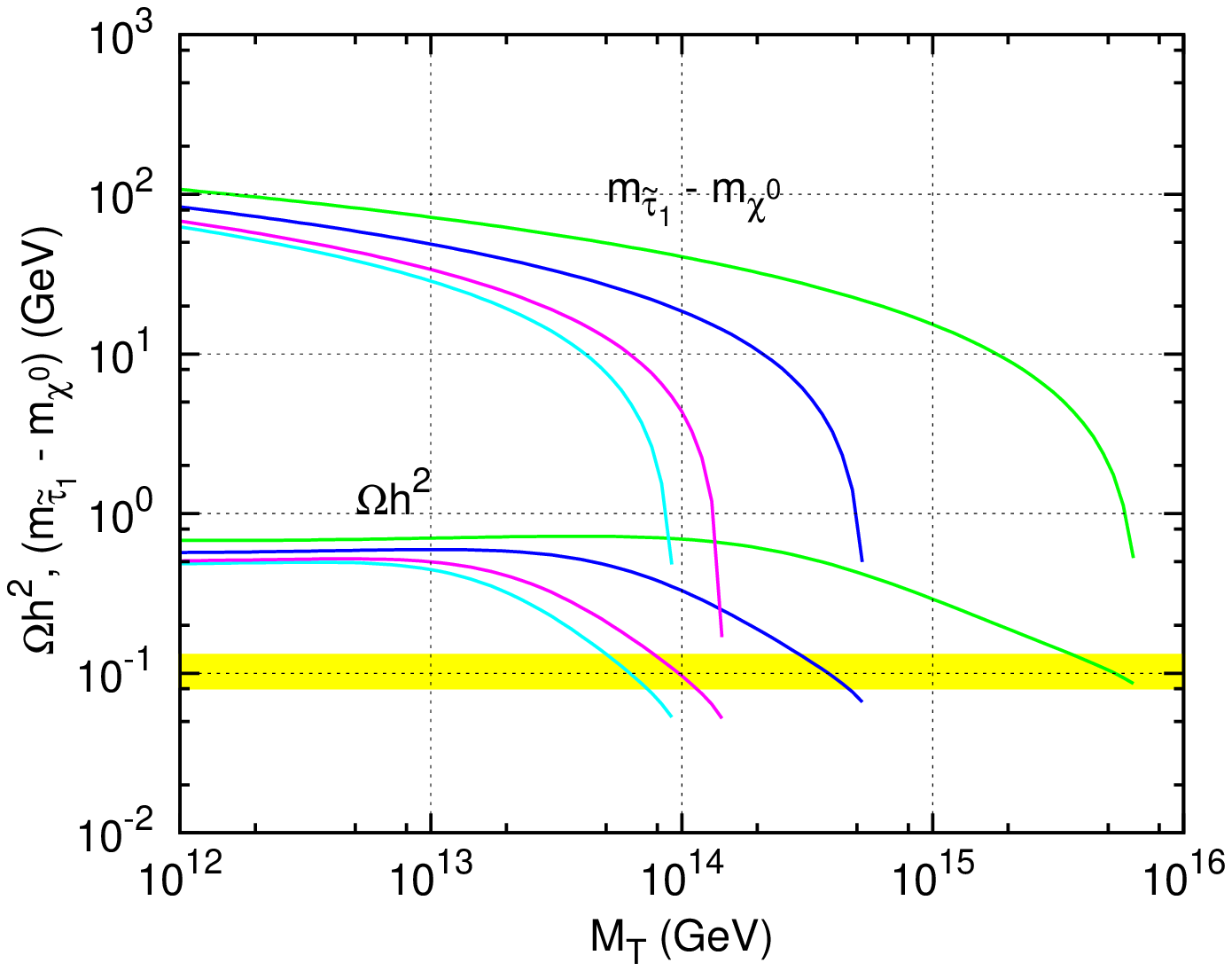}
\end{tabular}
\vspace{-5mm}
\caption{Allowed region for dark matter density
  ($0.081<\Omega_{\chi^0_1} h^2< 0.129$) 
in the ($m_0,M_{1/2}$) plane for the ``standard choice'' $\tan\beta=10$, 
$A_0=0$ and $\mu \ge 0$, for five values from $M_T$, $M_T=10^{14}$ GeV 
(red),  to  $M_T=10^{16}$ GeV (cyan), to the left. To the right: 
Variation of the mass difference $m_{\tilde{\tau}_1} -m_{\chi_0}$ 
(top lines) and of $\Omega h^2$ (bottom lines), as a function of $M_T$ 
for four different values of $m_0$: 0 (cyan), 50 (magenta), 100 (blue) 
and 150 GeV (green) for one fixed value of $M_{1/2}=800$ GeV. The yellow 
region corresponds to the experimentally allowed DM region. }
\label{fig:variationMT}
\end{figure}

Comparing the results for the pure mSugra case to the mSugra+15-plet 
calculation, two differences are immediately visible in fig. 
(\ref{fig:standard}). First, the focus point line is shifted towards 
larger values of $M_{1/2}$. This is due to the fact that for the 15-plet 
at $M_{\bf 15}=10^{14}$ GeV the neutralino is lighter than in the mSugra 
case at the same value of $M_{1/2}$, compare to fig. (\ref{fig:ana_para}). 
Maintaining the same relation between $M_1$ and $\mu$ as in the 
mSugra case requires a then a larger value of $M_{1/2}$. Note that for 
the same reason the excluded region from the LEP bound on the chargino 
mass is larger than in the mSugra case. Second on finds that the 
co-annihilation line is shifted towards smaller values of $m_0$. The 
latter can be understood from fig. (\ref{fig:variationMT}).

Fig. (\ref{fig:variationMT}) shows the allowed region for the dark matter 
density in the ($m_0,M_{1/2}$) plane for our ``standard choice'' of 
other mSugra parameters for a number of different $M_T$ (to the left). 
The plot shows how the co-annihilation line moves towards smaller values 
of $m_0$ for smaller values of $M_T$. The plot on the right in fig. 
(\ref{fig:variationMT}) explains this behaviour. It shows the variation 
of the mass difference $m_{\tilde{\tau}_1} -m_{\chi_0}$ (top lines) and 
of $\Omega h^2$ (bottom lines), as a function of $M_T$ for four different 
values of $m_0$: 0 (cyan), 50 (magenta), 100 (blue) and 150 GeV (green) 
for one fixed value of $M_{1/2}=800$ GeV. The yellow region corresponds 
to the experimentally allowed DM region. Co-annihilation requires 
a small value of $m_{\tilde{\tau}_1} -m_{\chi_0}$, typically smaller 
than a few GeV. With decreasing values of $M_T$ the gaugino masses 
run down to smaller values faster than the slepton masses, thus 
effectively increasing $m_{\tilde{\tau}_1} -m_{\chi_0}$ in these 
examples with respect to mSugra. To compensate for this effect at 
constant $M_{1/2}$ smaller values of $m_0$ are required to get the 
$m_{\tilde{\tau}_1} -m_{\chi_0}$ in the required  range.

At this point a short discussion of the Higgs boson mass bound might be in 
order. LEP excluded a light Higgs boson with SM couplings with masses 
below $m_{h} \le 114.4$ GeV \cite{PDG2008}. For reduced coupling 
of the Higgs boson to $b{\bar b}$ the bound is less severe, so this bound 
is not strictly valid in all of MSSM space. More important for us, however, 
is the {\em theoretical} uncertainty in the calculation of the lightest 
Higgs boson mass. SPheno calculates $m_{h^0}$ at two-loop level using 
$\overline{DR}$ renormalization. Expected errors for this kind of 
calculation, including a comparison of different public codes, have 
been discussed in \cite{Allanach:2004rh}. As discussed in 
\cite{Allanach:2004rh,Heinemeyer:2004xw} even at the 2-loop level 
uncertainties in the calculation of $m_{h^0}$ can be of the order of 
$3-5$ GeV. 
In this context it is interesting to note that FeynHiggs \cite{Frank:2006yh}, 
which calculates the higgs masses in a diagrammatic approach within 
the $\overline{OS}$ renormalization scheme tends to predict higgs 
masses which are systematically larger by $3-4$ GeV, when compared 
with the $\overline{DR}$ calculation. We therefore showed in fig. 
(\ref{fig:standard}) two lines of constant Higgs boson masses. The value 
of $m_{h^0}=114.4$ GeV is taking the LEP bound at face value, while 
the lower value of $m_{h^0}=110$ GeV estimates the parameter region 
which is excluded {\em conservatively}, including the theoretical 
error. Since the lightest Higgs boson mass varies slowly with $m_0$ and 
$M_{1/2}$, even a relatively tiny change in $m_{h^0}$ of, say 1 GeV, 
shifts the extreme values of the excluded region by $\sim 50$ GeV 
in $M_{1/2}$ (at small $m_0$) and by $\sim 150$ GeV in $m_0$ (at 
small $M_{1/2}$).

\begin{figure}[tb]
  \centering
  \begin{tabular}{cc}
   \includegraphics[height=60mm]{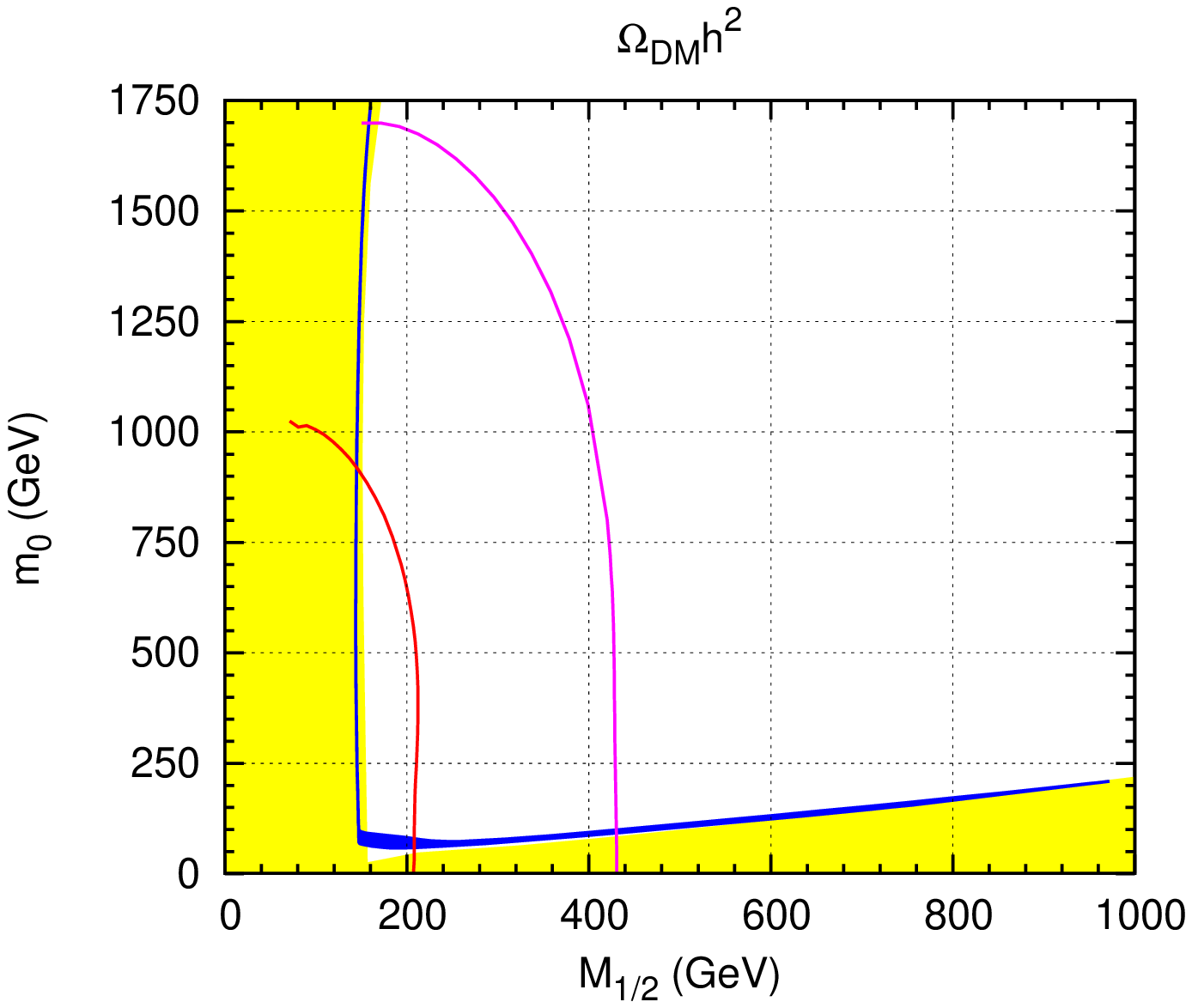}&
   \includegraphics[height=60mm]{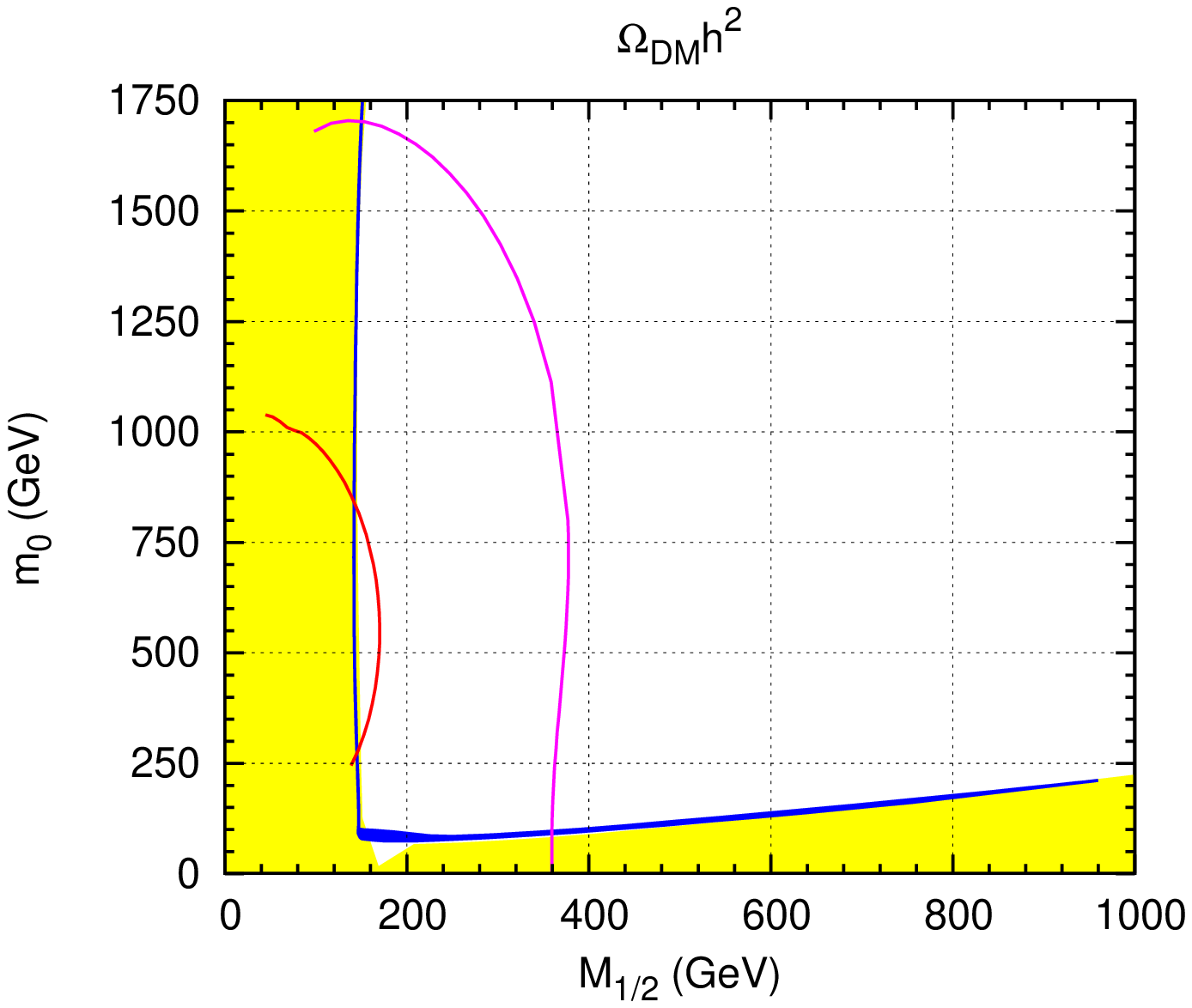}
  \end{tabular}
\caption{Limits for mSugra with  $\tan\beta=10$, and $\mu>0$ for 
$A_0=-300$ GeV (left panel) and $A_0=-500$ GeV (right panel). The 
blue regions are allowed by the DM constraint, for the explanation 
of the bounds see fig. (\ref{fig:standard}) and text.}
  \label{fig:HiggsLimitA0}
\end{figure}

Moreover, it is well known that the calculated Higgs boson masses are 
strongly dependent on the mixing in the stop sector and thus, 
indirectly, on the value of $A_0$. This is shown for the case of a 
pure mSugra calculation in fig. (\ref{fig:HiggsLimitA0}). Here 
we show two examples for the DM allowed region and the regions 
disfavoured by the Higgs boson mass bound at $m_{h^0}=114.4$ GeV and 
$m_{h^0}=110$ GeV. Larger negative $A_0$ leads to a less stringent 
constraint (for $\mu>0$). Note, that all of the bulk region becomes 
allowed at $A_0=-500$ GeV, once the theoretical uncertainty in 
the Higgs boson mass calculation is taken into account. We have checked 
for a few values of $M_T$ that for the case of mSugra+${\bf 15}$ 
the resulting Higgs boson bounds are very similar. We thus do not repeat 
the corresponding plots here. Comparing the calculations shown 
in fig. (\ref{fig:HiggsLimitA0}) and the mSugra calculation 
in fig. (\ref{fig:standard}) with each other, one finds that the 
DM allowed regions are actually affected very little by the choice 
of $A_0$. We have checked that this is also the case for 
mSugra + seesaw type-II.

\begin{figure}[!htb]
  \centering
  \begin{tabular}{cc}
\includegraphics[width=0.48\textwidth]{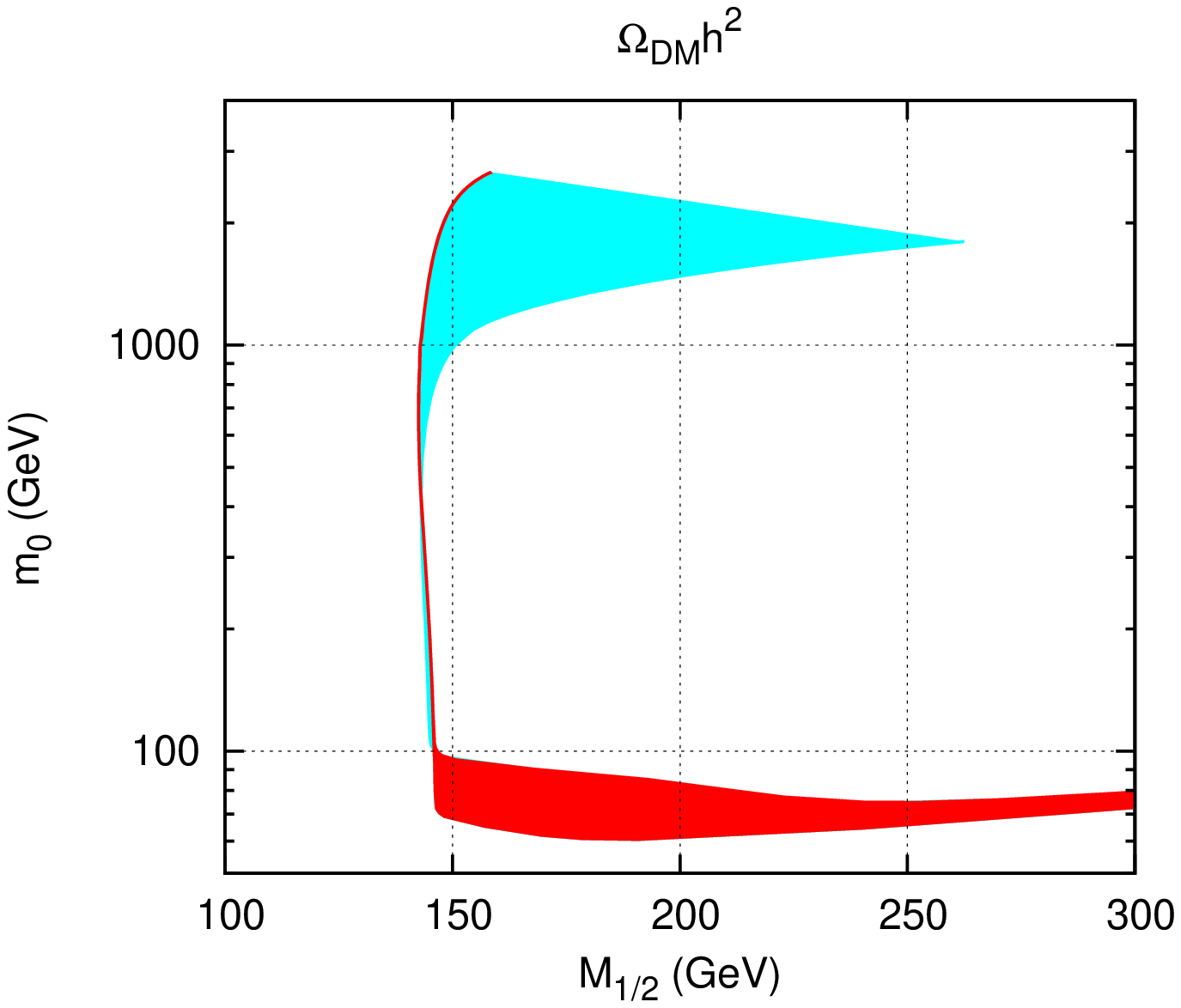}&
\includegraphics[width=0.48\textwidth]{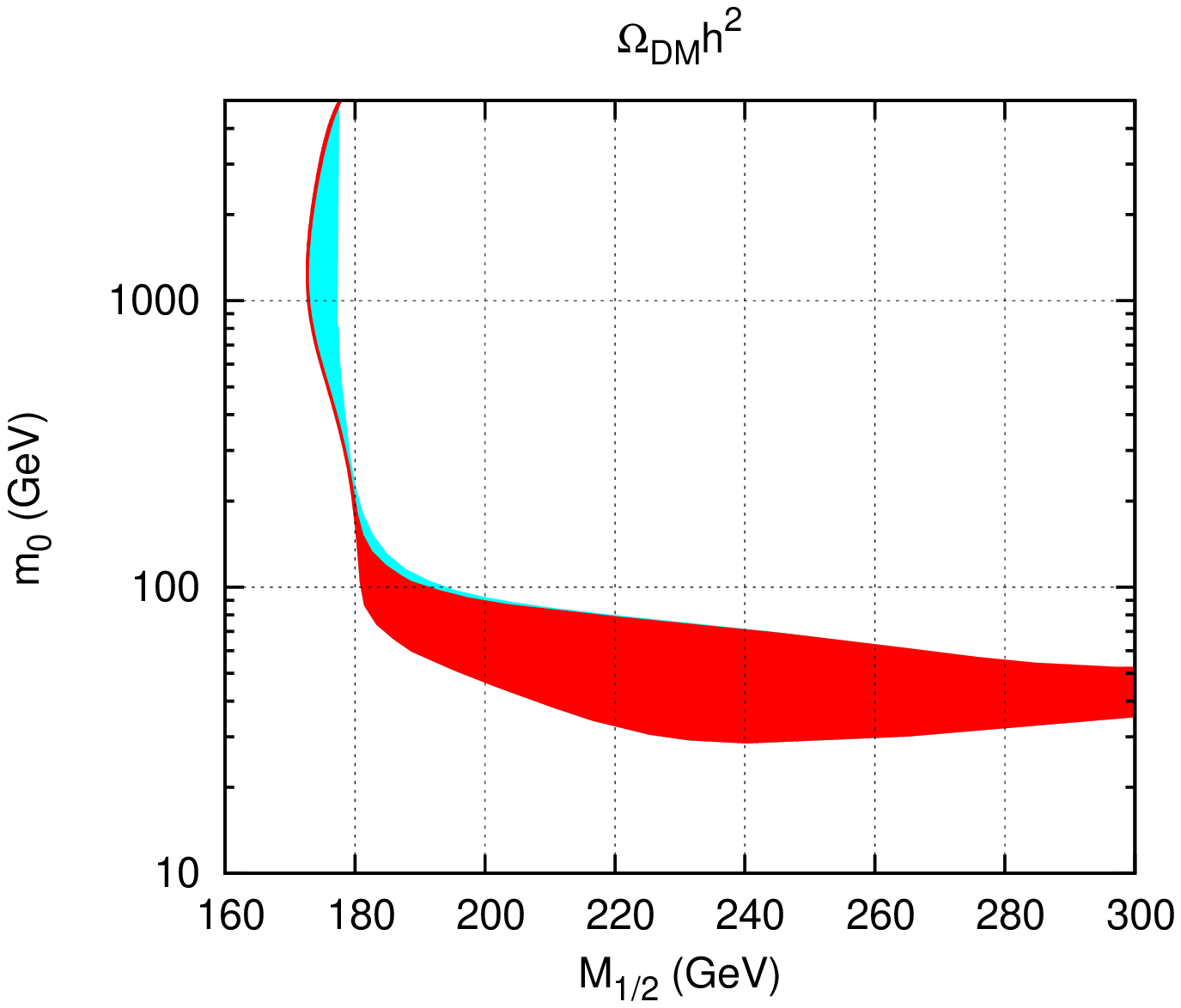}
\end{tabular}
\vspace{-5mm}
 \caption{Logarithmically scaled zoom into the focus point region. 
In red the allowed region for $0.081 < \Omega h^2 < 0.129$  and in 
cyan the allowed region due the variation of $m_{\rm top}=171.2 \pm 2.1$ 
GeV. The left panel is for mSugra case and the right panel for $M_T=10^{15}$ 
GeV. The other parameters are taken at our ``standard'' values.}
\label{fig:mtop0}
\end{figure}

As mentioned above the uncertainty in the top mass is important for 
the calculation of the relic density. At low and moderate values of 
$\tan\beta$ the exact value of $m_t$ affects mainly the focus point 
region. As fig. (\ref{fig:standard}) demonstrates near the focus point 
line the relic density changes very abruptly even for tiny changes of 
$M_{1/2}$. This is because a comparatively small value of $\mu$ is 
required to get a sufficiently enhanced coupling of the neutralino 
to the $Z^0$ boson. In mSugra the value of $\mu$ is determined from 
all other parameters by the condition of having correct electro-weak 
symmetry breaking (EWSB) and usually leads to $M_1,M_2 \ll \mu$. In 
the focus point region $\mu$ varies abruptly, points to the ``left'' 
of the focus point region are usually ruled out by the fact that EWSB 
can not be achieved. Since $m_t$ is the largest fermion mass, its 
exact value influences the value of $\mu$ required to achieve EWSB most. 
The change of $\mu$ with respect to a change of $m_t$ then can lead 
to a significant shift in the DM allowed region of parameter space. 
This is demonstrated in fig. (\ref{fig:mtop0}), which shows a zoom 
into the focus point region for pure mSugra (to the left) and mSugra + 
${\bf 15}$ (to the right). The variation of the top mass shown corresponds 
to the current 1 $\sigma$ allowed range \cite{PDG2008}. The pure mSugra 
is especially sensitive to a change of $m_t$. At large values of $m_0$ 
the uncertainty in ``fixing'' $M_{1/2}$ from the DM constraint can 
be larger than $100$ GeV in the case of mSugra. Given this large 
uncertainty it would be impossible at present to distinguish 
the pure mSugra case from mSugra + seesaw, if the focus point region 
is the correct explanation of the observed DM. Note, however, that 
in the future the top mass will be measured more precisely. At 
the LHC one expects an uncertainty of 1-2 GeV \cite{Aad:2009wy} 
at a linear collider $m_t$ could be determined down to an uncertainyy of 
$100$ MeV \cite{AguilarSaavedra:2001rg}.

\begin{figure}[!htb]
  \centering
\includegraphics[width=0.6\textwidth]{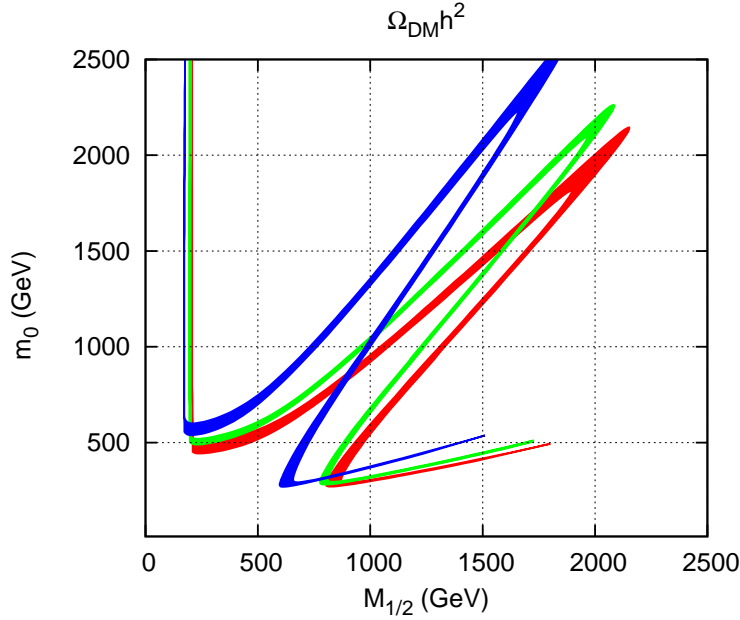}
\vspace{-5mm}
\caption{Allowed region for dark matter density in the ($m_0,M_{1/2}$) 
plane for $A_0=0$,  $\mu \ge 0$ and  $\tan\beta=45$, for (from top to 
bottom) $M_T=5\times 10^{13}$ GeV (red), $M_T=10^{14}$(green) and 
$M_T=10^{15}$ GeV (blue).}
\label{fig:funMT}
\end{figure}

We now turn to a discussion of large $\tan\beta$. At large values of 
$\tan\beta$ the width of the CP-odd Higgs boson $A$ becomes large, 
$\Gamma_A \sim M_A \tan^2\beta (m_b^2+m_t^2)$, and a wide s-channel 
resonance occurs in the region $m_{\chi^0_1} \simeq M_A/2$. 
The enhanced annihilation cross section reduces $\Omega_{\chi^0_1}h^2$ 
to acceptable levels, the resulting region is known as the ``higgs 
funnel'' region. In fig. (\ref{fig:funMT}) we show the allowed 
range of parameters in the ($m_0,M_{1/2}$) plane for one specific 
value of $\tan\beta=45$ and three different values of $M_T$. As 
demonstrated, the higgs funnel region is very sensitive to the 
choice of $M_T$. It is fairly obvious that varying $M_T$ one can 
cover nearly all of the plane, even for fixed values of all other 
parameters. We have calculated the DM allowed region for various 
values of $\tan\beta$ and found that the funnel appears for all 
$\tan\beta\gsim 40$, approximately.

\begin{figure}[!htb]
  \centering
\includegraphics[width=0.45\textwidth]{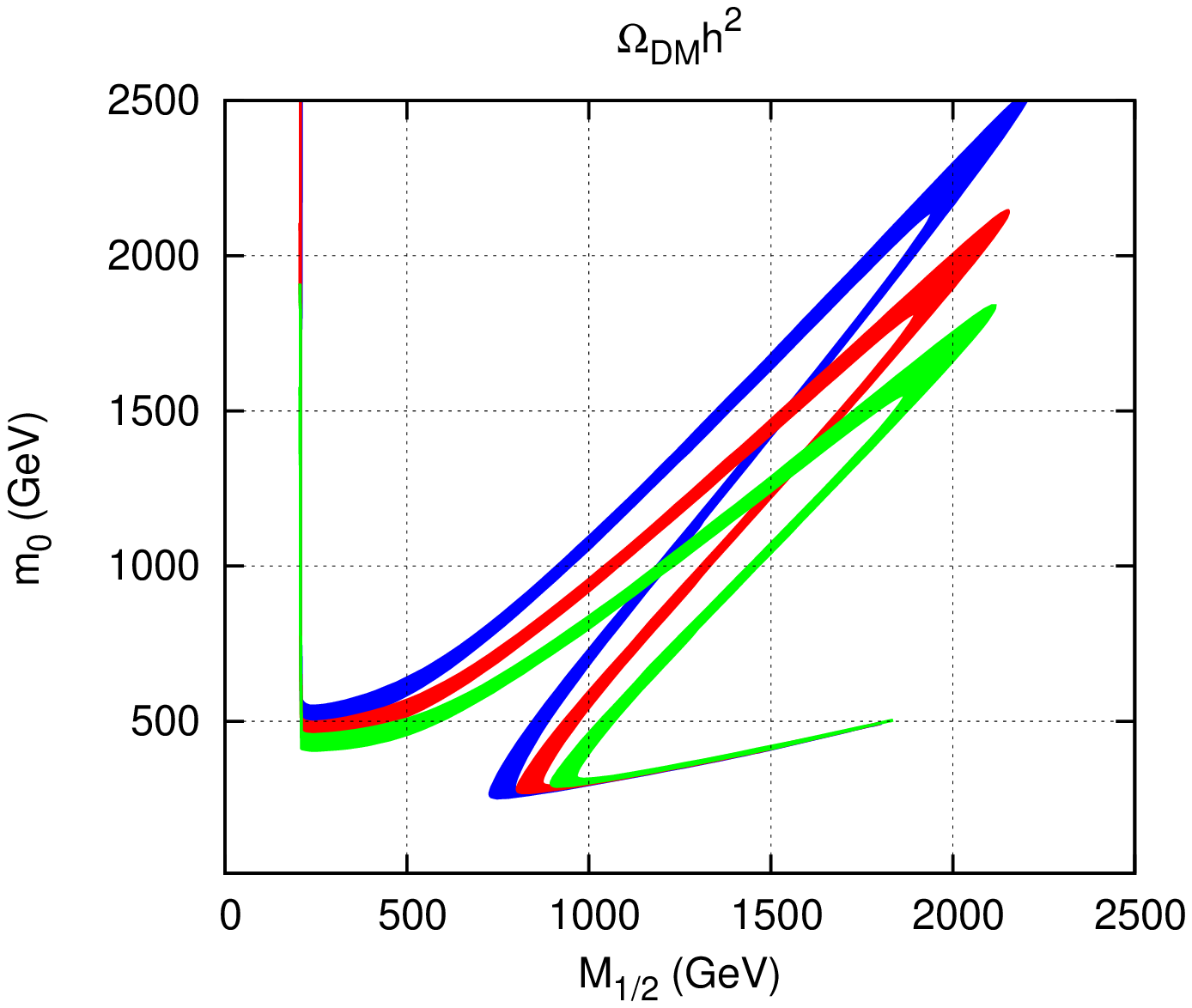}
\includegraphics[width=0.45\textwidth]{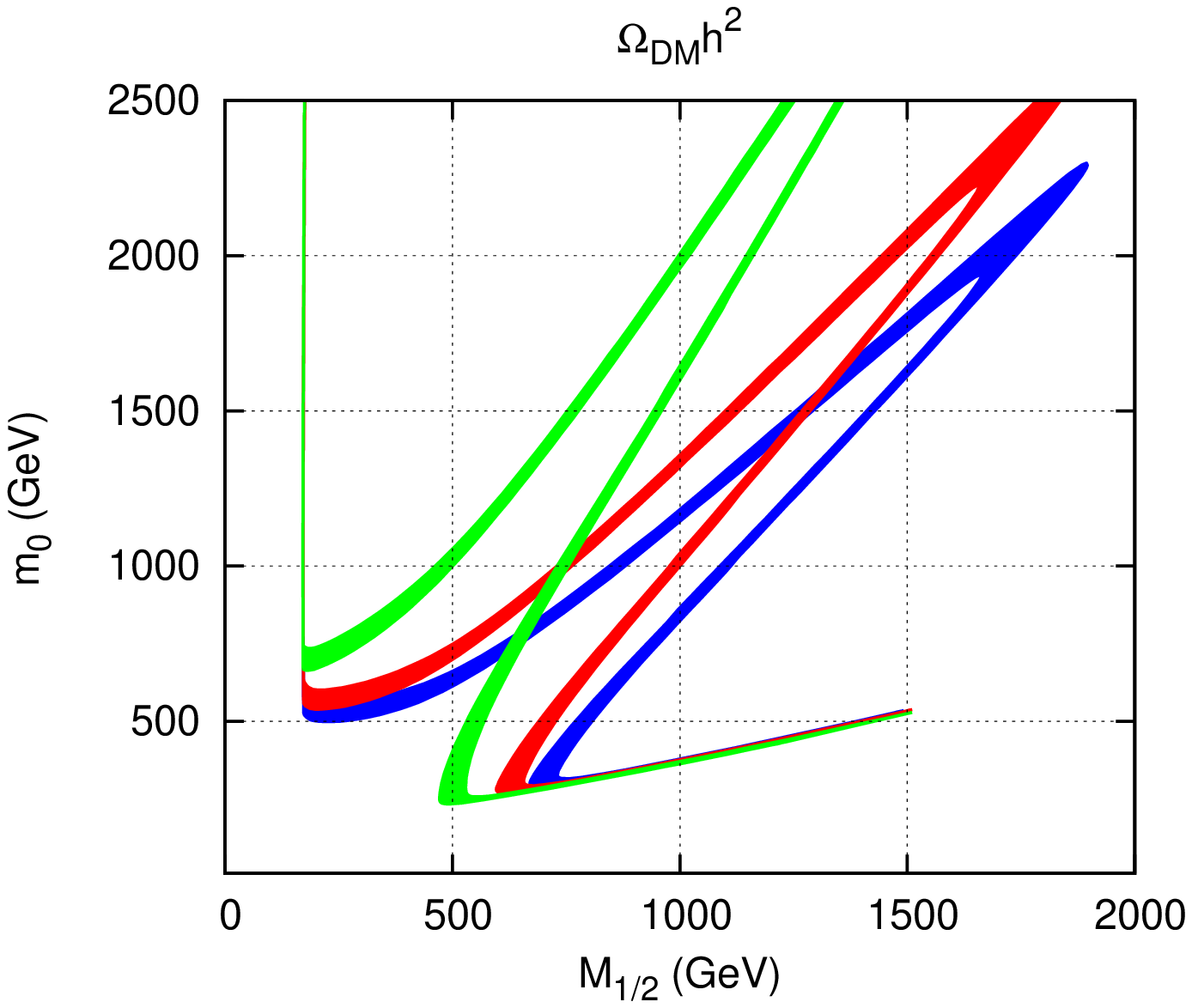}
\vspace{-5mm}
\caption{Allowed region for the dark matter density in the ($m_0,M_{1/2}$) 
plane for $A_0=0$,  $\mu \ge 0$ and  $\tan\beta=45$, for 
$M_T=5\times 10^{13}$ GeV and (to the left) for three values of 
$m_{top}=169.1$GeV (blue), $m_{top}=171.2$ GeV (red) and 
$m_{top}=173.3$ GeV (green). To the right: The same, but varying $m_b$. 
$m_{bot}=4.13$ GeV (blue), $m_{bot}=4.2$ GeV (red) and $m_{bot}=4.37$ 
GeV (green).}
\label{fig:mtmb}
\end{figure}

The strong dependence of the higgs funnel region on $M_T$ unfortunately 
does not imply automatically that if large $\tan\beta$ is realized 
in nature one could get a very sensitive indirect ``measurement'' of 
the seesaw scale by determining ($m_0,M_{1/2}$). The reason is that  
the higgs funnel is also very sensitive to the exact value of 
$\tan\beta$ and to the values (and errors) of the top and bottom 
quark mass. The latter is demonstrated in fig. (\ref{fig:mtmb}), where 
we show the DM allowed range of parameters for a fixed choice of 
$\tan\beta$ and $M_T$ varying to the left (to the right) $m_t$ ($m_b$) 
within their current 1 $\sigma$ c.l. error band. The position of the 
funnel is especially sensitive to the exact value of $m_b$. Comparing 
fig. (\ref{fig:mtmb}) with fig. (\ref{fig:funMT}) one can see that 
the uncertainty in $m_b$ and $m_t$ currently severely limit any 
sensitivity one could get on $M_T$. However, future determinations 
of $m_b$ and $m_t$ could improve the situation considerably. For 
future uncertainties in $m_t$ see the discussion above for the 
focus point region. For $m_b$ reference \cite{Brambilla:2004wf} 
estimates that $m_b$ could be fixed to $4.17\pm 0.05$ GeV, 
which might even be improved to an accuracy of $\Delta m_b\simeq 16$ MeV 
according to \cite{Chetyrkin:2009fv}.

\begin{figure}[!htb]
  \centering
  \begin{tabular}{cc}
\includegraphics[width=0.48\textwidth]{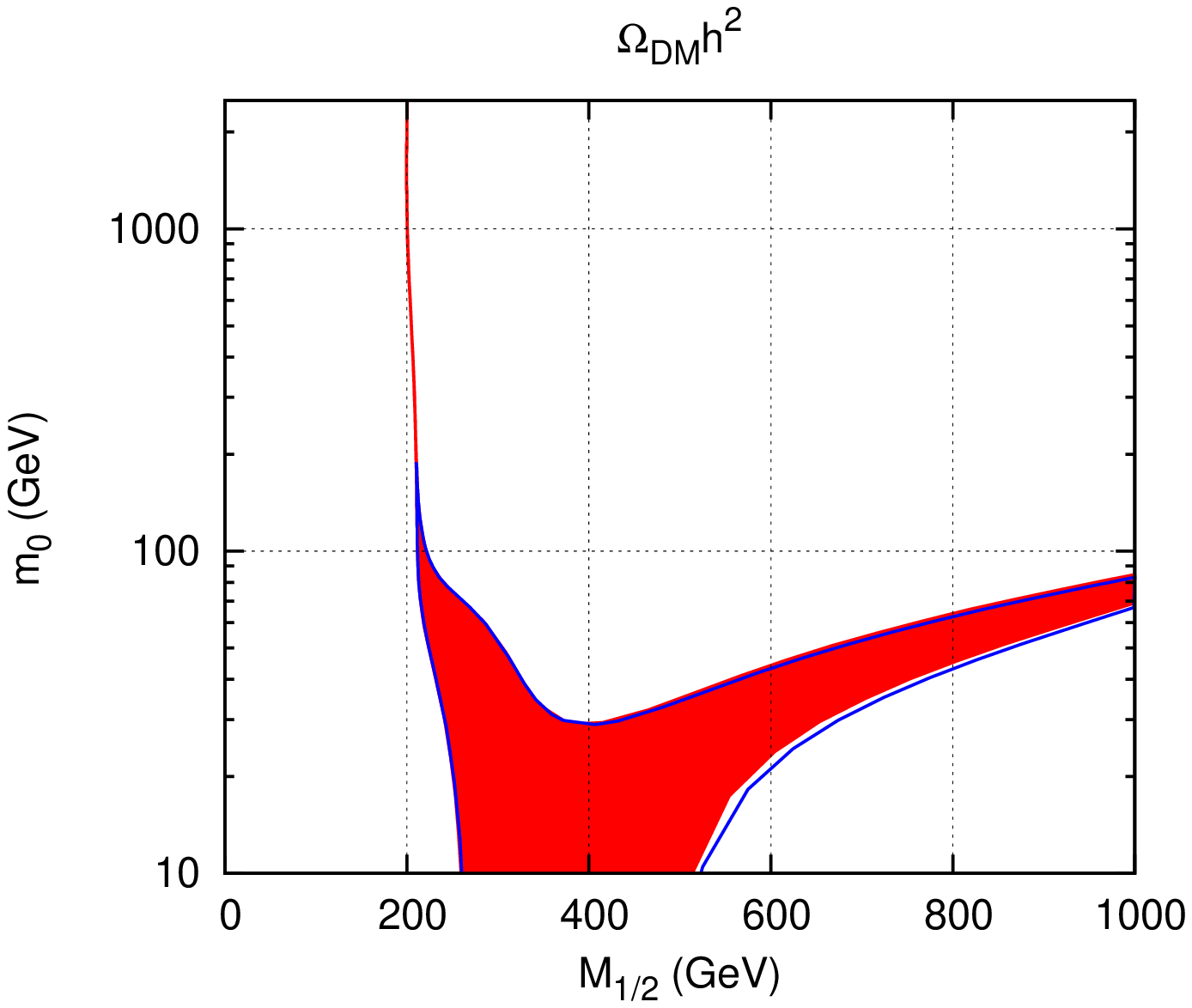}
\includegraphics[width=0.48\textwidth]{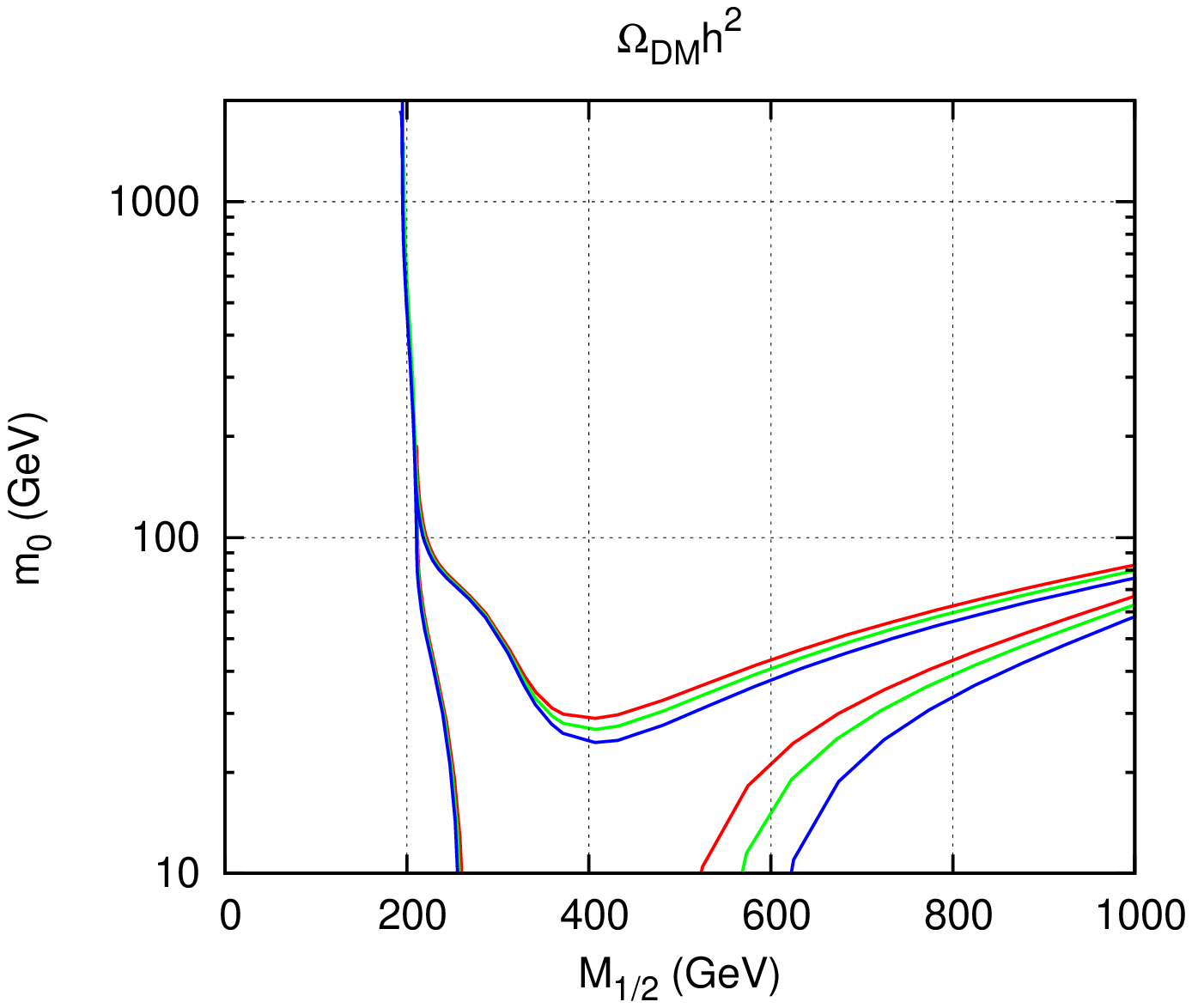}
\end{tabular}
\vspace{-5mm}
\caption{Allowed region for dark matter density
in the ($m_0,M_{1/2}$) plane for the ``standard choice'' of mSugra 
parameters for $M_T=10^{14}$ GeV. To the left: For one fixed value 
of $\lambda_2=0.5$ the allowed range for negligibly small neutrino 
Yukawa couplings (red) and $Y_T$ fitted to correctly explain solar 
and atmospheric neutrino data (blue lines). To the right: the 
DM allowed range of parameters for 3 different values of $\lambda_2$, 
$\lambda_2=0.5$ (red), $\lambda_2=0.75$ (green) and $\lambda_2=1$ (blue). 
Note the logarithmic scale.}
\label{fig:nufit}
\end{figure}

All of the above figures have been calculated using fixed values for 
$\lambda_1$ and $\lambda_2$ and negligibly small Yukawa couplings 
$Y_T$. This choice in general does not affect the calculation of 
the DM allowed regions much. However, a fully consistent calculation 
can not vary $M_T$, $Y_T$ and $\lambda_2$ independently, since this 
will lead to neutrino masses and angles outside the experimentally 
allowed ranges. Since $Y_T$ is diagonalized by the same matrix as 
the effective neutrino mass matrix, $m_{\nu}$, see the previous 
section, the measured neutrino angles provide constraints on the 
relative size of the entries in $Y_T$. The absolute size of $Y_T$ is 
then fixed for any fixed choice of $\lambda_2$ and $M_T$, once the 
neutrino spectrum is chosen to be hierarchical or quasi-degenerate. 
In the numerical calculation shown in fig. (\ref{fig:nufit}) we have 
chosen neutrino masses to be of the normal hierarchical type and 
fitted the neutrino angles to exact tri-bimaximal (TBM) values 
\cite{Harrison:2002er}, i.e. $\tan^2\theta_{\rm Atm}=1$, 
$\tan^2\theta_{\odot}=1/2$ and $\sin^2\theta_{\rm R}=0$. This 
has to be done in a simple iterative procedure, since the triplet 
parameters are defined at the high scale, whereas neutrino masses 
and angles are measured at low scale. For more details on the 
fit procedure see \cite{Hirsch:2008gh}.

In fig. (\ref{fig:nufit}) to the left we show two calculations of 
the DM allowed regions. The allowed range for negligibly small neutrino 
Yukawa couplings is shown by the filled (red) region, while the 
calculation with $Y_T$ fitted to correctly explain solar 
and atmospheric neutrino data is the one inside the (blue) lines.
Note the logarithmic scale. As demonstrated, the exact values of 
$Y_T$ are of minor importance for the determination of the parameter 
region allowed by the DM constraint. Slightly larger differences 
between the fitted and unfitted calculations are found pushing 
$M_T$ to larger values (see, however, below). For smaller values of 
$M_T$, the entries in $Y_T$ needed to correctly explain neutrino data 
are smaller and, thus, $Y_T$ affects the DM allowed region 
even less for $M_T < 10^{14}$ GeV.

In fig. (\ref{fig:nufit}) to the right we compare three different 
calculations for $\lambda_2$, $\lambda_2=0.5$ (red), $\lambda_2=0.75$ 
(green) and $\lambda_2=1$ (blue), for fixed choice of other parameters. 
This plot serves to show that also the exact choice of $\lambda_2$ 
is of rather minor importance for the determination of the DM 
allowed region. Very similar results have been found for $\lambda_1$, 
we therefore do not repeat plots varying $\lambda_1$ here.

\begin{figure}[!htb]
  \centering
  \begin{tabular}{cc}
\includegraphics[width=0.48\textwidth]{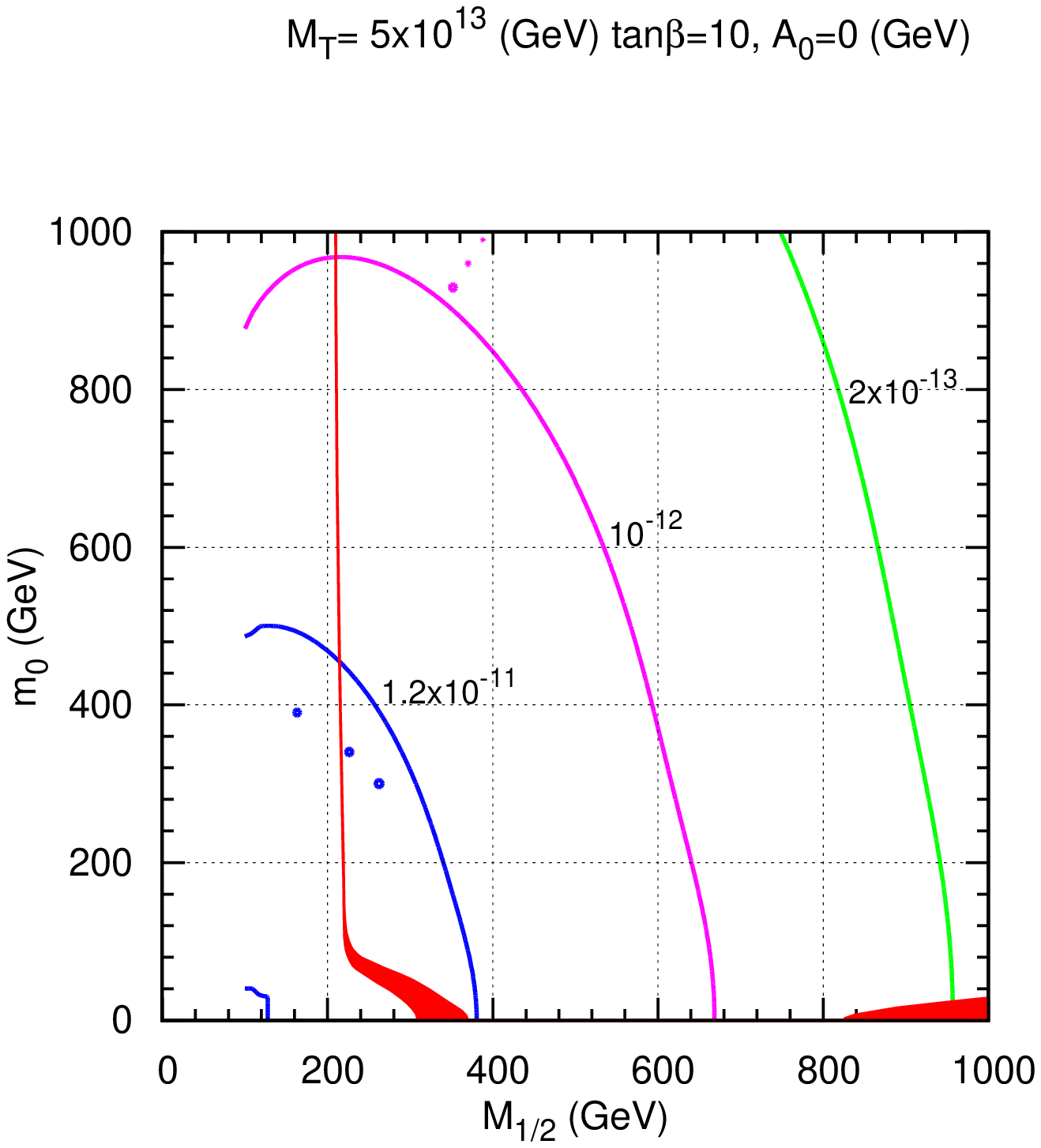}
\includegraphics[width=0.48\textwidth]{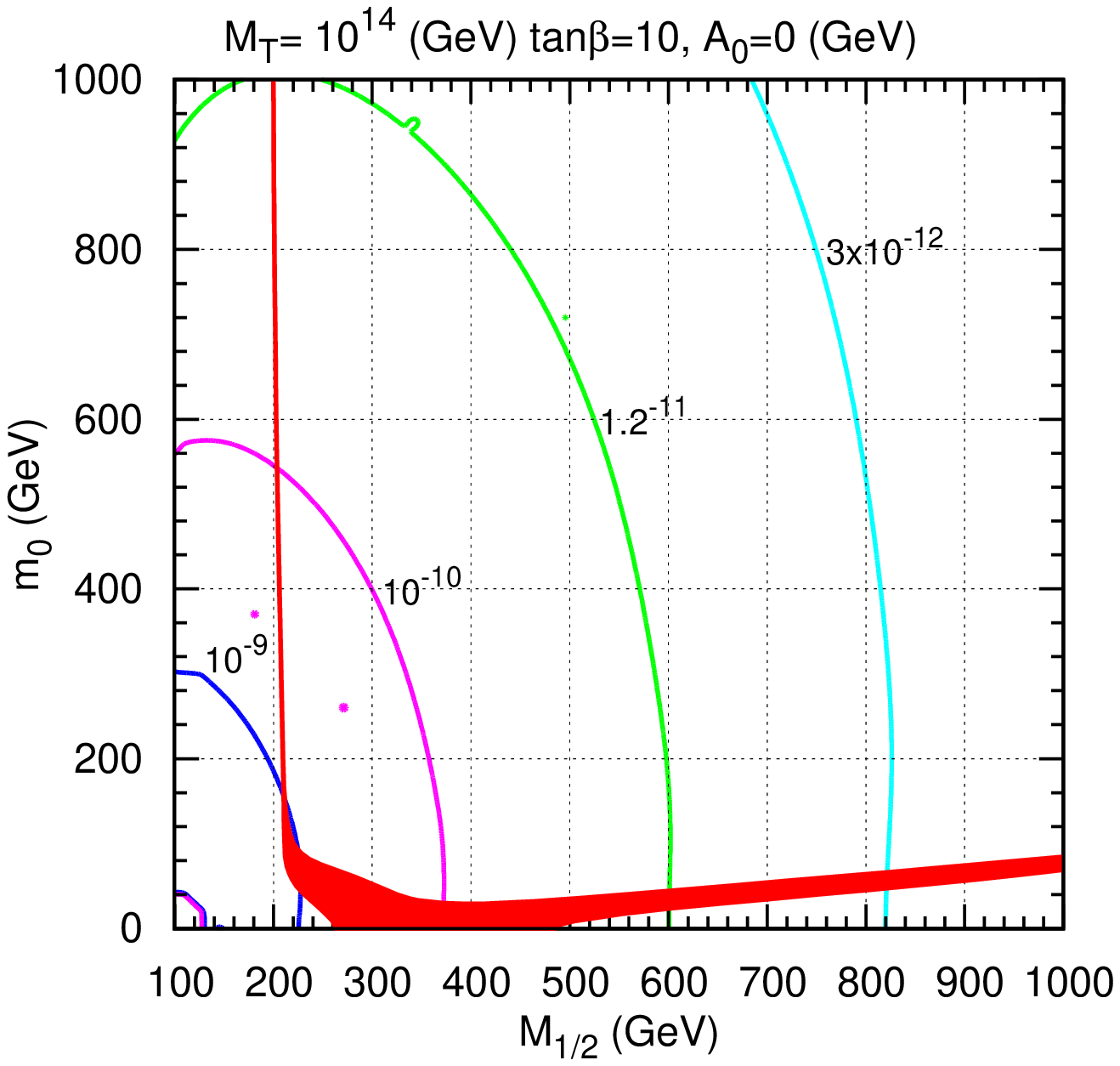}
\end{tabular}
\vspace{-5mm}
\caption{Allowed region for dark matter density 
in the ($m_0,M_{1/2}$) plane for our ``standard choice'' of mSugra 
parameters and for two values of $M_T$: $M_T=5\times 10^{13}$ (left 
panel) and for $M_T=10^{14}$ (right panel). Superimposed are the contour 
lines for the $Br(\mu\rightarrow e \gamma)$.}
\label{fig:LFVtb10}
\end{figure}

Finally, we will compare the constraints imposed on the parameter 
space of the model by $\Omega_{DM}h^2$ with the constraints from 
the current data on non-observation of lepton flavour violating 
processes. Since LFV  within the present model has been studied 
in some detail in \cite{Hirsch:2008gh}, we will not repeat all of 
the discussion here. Instead, here we concentrate on $\mu\to e \gamma$ 
exclusively, since the upper bound on $Br(\mu\rightarrow e \gamma)$ 
of $Br(\mu\rightarrow e \gamma)\le 1.2 \cdot 10^{-11}$ 
\cite{PDG2008} has been shown to provide currently the most important 
constraint. 

In fig. (\ref{fig:LFVtb10}) we show the DM allowed parameter regions 
for $\tan\beta=10$ and two values of $M_T$, $M_T=5\cdot 10^{13}$ GeV 
(to the left) and $M_T=10^{14}$ GeV (to the right), for a fixed 
choice of all other parameters. Superimposed on this plot are lines 
of constant branching ratio for $Br(\mu\rightarrow e \gamma)$. The 
latter have been calculated requiring neutrino masses being hierarchical 
and fitted to solar and atmospheric neutrino mass squared differences 
and neutrino angles fitted to TBM values. Within the ($m_0,M_{1/2}$) 
region shown, $Br(\mu\rightarrow e \gamma)$ can vary by two 
orders of magnitude, depending on the exact combination of ($m_0,M_{1/2}$), 
even for all other parameters fixed. The most important parameter 
determining $Br(\mu\rightarrow e \gamma)$, once neutrino data is 
fixed, however, is $M_T$, as can be seen comparing the figure to 
the left with the plot on the right. While for $M_T=10^{14}$ GeV 
about ``half'' of the plane is ruled out by the non-observation 
of $\mu\to e \gamma$, for $M_T=5\cdot 10^{13}$ GeV with the 
current upper limit nearly all of the plane becomes allowed. The strong 
dependence of $\mu\to e \gamma$ on $M_T$ can be understood from the 
analytical formulas presented in \cite{Hirsch:2008gh}. In this paper 
it was shown that $Br(\mu\rightarrow e \gamma)$ scales very 
roughly as $Br(\mu\rightarrow e \gamma)\propto M_T^4 \log(M_T)$, 
if neutrino masses are to be explained correctly. For $\tan\beta=10$ 
one thus concludes that with present data values of $M_T$ larger 
than (few) $10^{13}$ GeV - (few) $10^{14}$ GeV are excluded by 
$Br(\mu\rightarrow e \gamma)$, to be compared with $M_T/\lambda_2 \lsim 
10^{15}$ GeV from the measured neutrino masses. Note, however, that 
(i) the constraint from neutrino masses is relatively independent of 
$\tan\beta$, $m_0$ and $M_{1/2}$, while $\mu\to e \gamma$ shows strong 
dependence on these parameters; and (ii) allowing the value of the reactor 
angle $\sin^2\theta_{\rm R}$ to vary up to its experimental upper limit, 
$\sin^2\theta_{\rm R}=0.056$ \cite{Maltoni:2004ei}, leads to larger 
values of $Br(\mu\rightarrow e \gamma)$ and thus to a tighter 
upper limit on $M_T$. 

\begin{figure}[!htb]
  \centering
  \begin{tabular}{cc}
\includegraphics[width=0.45\textwidth]{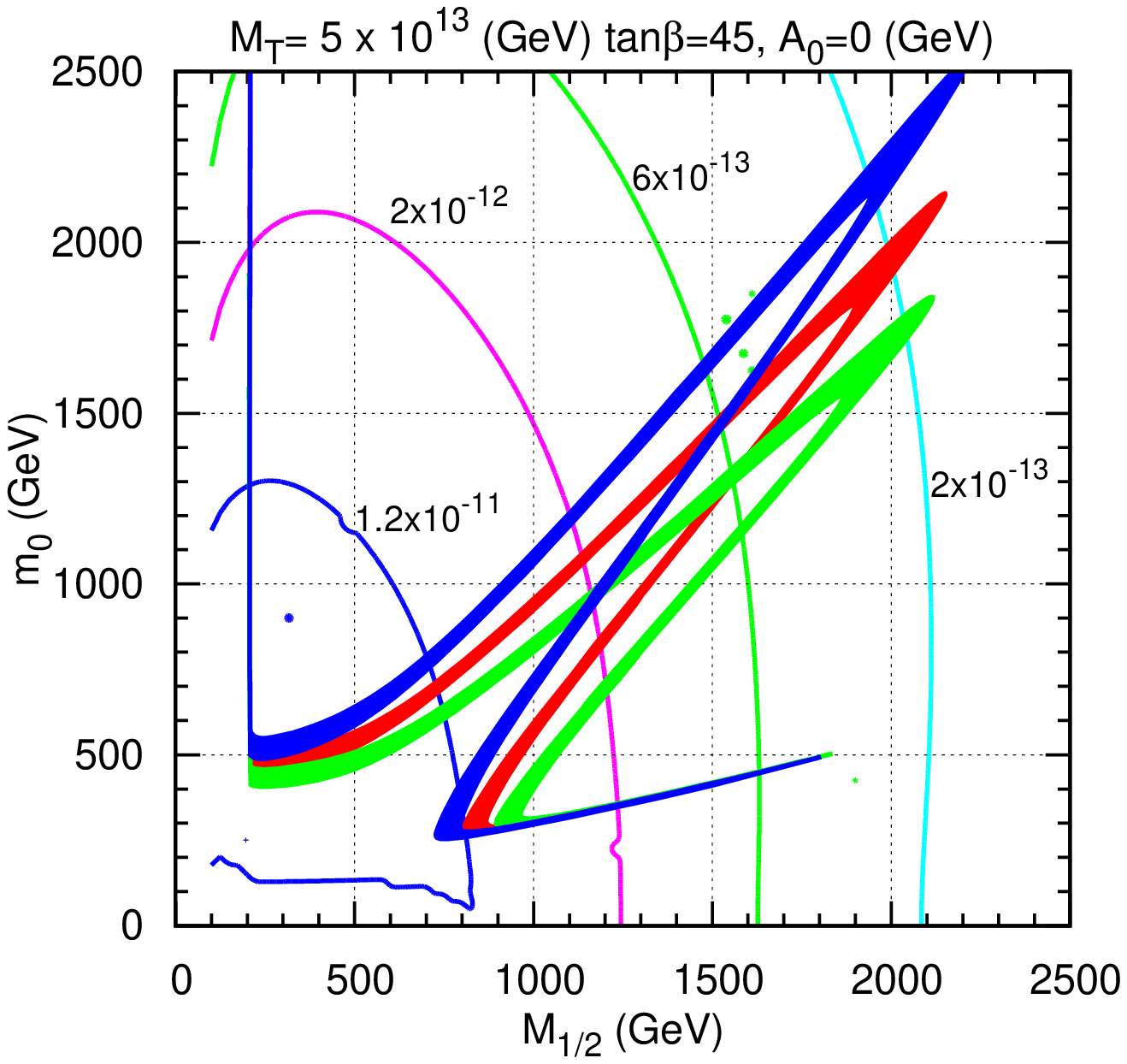}&
\includegraphics[width=0.45\textwidth]{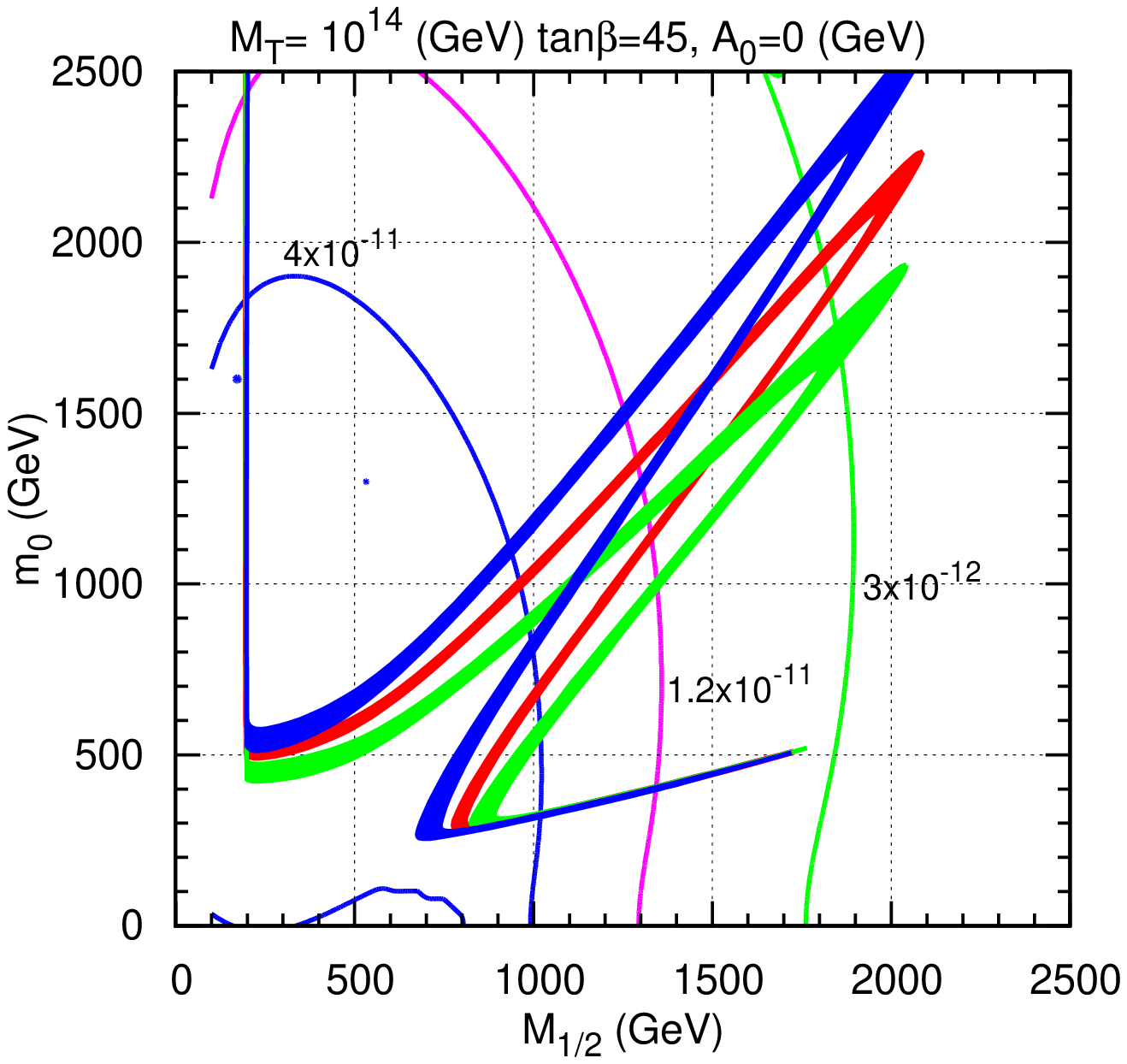}
  \end{tabular}
\vspace{-5mm}
\caption{Allowed region for dark matter density
  ($0.081<\Omega_{\chi^0_1} h^2< 0.129$) 
in the ($m_0,M_{1/2}$) plane for $A_0=0$,  $\mu \ge 0$ and
$\tan\beta=45$, for three values of $m_{top}=169.1$
GeV (blue), $m_{top}=171.2$ GeV (red) and $m_{top}=173.3$ GeV (green)
for $M_T=5\times 10^{13}$ (left panel) and for $M_T=10^{14}$ (right
panel). Superimposed are the contour lines for the $Br(\mu\rightarrow
e \gamma)$.}
\label{fig:LFVtb45}
\end{figure}

In fig. (\ref{fig:LFVtb45}) we show the results for a calculation 
comparing dark matter and LFV in the case of large $\tan\beta$. Here 
the same constraints as in fig. (\ref{fig:LFVtb10}) are shown, however 
for $\tan\beta=45$. Again we show the calculation for two values of 
$M_T$, since $M_T$ is the most important free parameter. It is known 
that at large values of $\tan\beta$, LFV decays are enhanced due 
to an enhanced chargino diagram, which in 
the limit of large $\tan\beta$ scales approximately as $\tan^2\beta$~\cite{Hisano:1995cp}. 
Therefore, constraints on the parameter space from non-observation of 
LFV decays are more severe in case of large $\tan\beta$, leading to 
tighter upper limits on $M_T$. This is clear if we compare
fig. (\ref{fig:LFVtb10}) and fig. (\ref{fig:LFVtb45}), noticing the
different scales. However, because of the higgs funnel region
developing for large $\tan\beta$, the interesting part of the parameter
space enlarges compensating for the larger values of the LFV decays. 
This can be seen in fig. (\ref{fig:LFVtb45}), where for
$M_T=5\times10^{13}$ GeV (left), most of the ($m_0,M_{1/2}$) plane is 
allowed by the upper limit on $Br(\mu\rightarrow e \gamma)$, while 
for $M_T =10^{14}$ GeV (right),  about ``half'' of the plane is ruled out 
by this limit.

We have concentrated in this paper on discussing DM in mSugra with a 
seesaw type-II. Before closing this section, we would like to briefly 
comment on the case of seesaw type-I. In seesaw type-I one adds two 
or more singlet superfields to the superpotential of the MSSM. These 
singlets have Yukawa couplings to the standard model lepton doublet 
and a Majorana mass term, but no other couplings to any of the MSSM 
fields. The running of the mSugra soft parameters in this setup is 
therefore only changed by the neutrino Yukawa couplings. Just as in 
the seesaw type-II one can estimate from current neutrino data that 
the Yukawa couplings are order $Y^{\nu} \sim {\cal O}(1)$ for the 
right-handed Majorana mass order ${\cal O}(10^{15})$ GeV. 
\footnote{Different from the seesaw type-II, where $Y_T$ depends linearly 
on $M_T$, however, in seesaw type-I Yukawas scale like $Y^{\nu} \sim 
\sqrt{M_M}$.}
For any $M_M$ smaller than this number, one therefore expects that 
the running of the soft parameters is essentially mSugra-like. 
(Apart from small off-diagonal terms in $m_{L}^2$, which are exactly 
zero in mSugra.) This implies that also the DM regions should be very 
close to those found in the mSugra case. We have confirmed this expectation 
by calculating the DM allowed region for our standard choice of 
mSugra parameters and various values of the right-handed neutrino masses. 
Even for $Y^{\nu}$ at the upper limit allowed by perturbativity 
we did not find any significant departure from the mSugra case. With 
the hindsight of the results shown in fig. (\ref{fig:nufit}) for the 
seesaw type-II this is not surprising.

One exceptional case for the seesaw type-I has been discussed, however, 
recently in \cite{Kadota:2009vq}. The authors of \cite{Kadota:2009vq} 
observed that for Yukawa couplings close to one and a large value of 
the common trilinear $A_0$, say $A_0 = 1100$ GeV, the left sneutrinos 
can be the next-to-LSP (NLSP) for small-to-moderate values of 
$m_0$, $M_{1/2}$ and $\tan\beta$. For a sneutrino NLSP nearly 
degenerate with the lightest neutralino a new co-annihilation 
regions then shows up at small values of $m_0$. We have repeated 
this calculation with three right-handed neutrinos (\cite{Kadota:2009vq} 
use only one singlet superfield) and confirm the sneutrino 
co-annihilation region for $|Y^{\nu}| \simeq {\cal O}(1)$ and large $A_0$. 
However, in our calculation, if we insist on fitting the large atmospheric 
and solar angles, all of the region is excluded by upper limits on 
LFV decays, if we put the matrix $R$ of the Casas-Ibarra parameterization 
for the neutrino Yukawa couplings \cite{Casas:2001sr} to the identity 
matrix. As has been shown in \cite{Ellis:2002fe,Arganda:2005ji}, we 
could, in principle, avoid these strong constraints from LFV by 
a careful adjustments of the unknown parameters in $R$. We did, 
however, not attempt to do a systematic study as to how $R$ has 
to be chosen that the sneutrino co-annihilation becomes consistent 
with LFV decays.

\section{Conclusions}
\label{sec:cncl}

In conclusion, we have calculated the neutralino relic density in a 
supersymmetric model with mSugra boundary conditions including a 
type-II seesaw mechanism to explain current neutrino data. We have 
discussed how the allowed ranges in mSugra parameter space change as 
a function of the seesaw scale. The stau co-annihilation region is 
shifted towards smaller $m_0$ for smaller values of the triplet mass 
$M_T$, while the bulk region and the focus point line are shifted 
towards larger values of $M_{1/2}$ for $M_T$ sufficiently below the 
GUT scale. The higgs funnel, which appears at large values of $\tan\beta$ 
has turned out to be especially sensitive to the value of $M_T$. 
Determining $M_{1/2}$ from the mass of any gaugino and $m_0$ from 
a sparticle which is {\em not important for the DM calculation}, 
one could, therefore, get a constraint on $M_T$ from the requirement 
that the observed $\Omega_{DM}h^2$ is correctly explained by the 
calculated $\Omega_{\chi^0_1}h^2$.

On the positive side, we can remark that current data on neutrino masses 
put an upper bound on $M_T$ of the order of ${\cal O}(10^{15})$ GeV. 
Since this is at least one order of magnitude smaller than the GUT 
scale, the characteristic shifts in the DM regions are necessarily 
non-zero if our setup is the correct explanation of the observed 
neutrino oscillation data. Even more stringent upper limits on $M_T$ 
follow, in principle, from the non-observation of LFV decays. A 
smaller $M_T$ implies larger shifts of the DM region. However, the 
``exact'' upper limit on $M_T$ from LFV decays depends strongly on 
$\tan\beta$, $m_0$ and $M_{1/2}$, and thus can be quantified only 
once at least some information on these parameters is available. 

On the down side, we need to add a word of caution. We have found that 
the DM calculation suffers from a number of uncertainties, even if 
we assume the soft masses to be perfectly known. The most important 
SM parameters turn out to be the bottom and the top quark mass. 
The focus point line depends extremely sensitively on the exact value 
of the top mass, the higgs funnel shows a strong sensitivity on both, 
$m_b$ and $m_t$.

Finally, it is clear that quite accurate sparticle mass measurements 
will be necessary, before any quantitative conclusions can be taken 
from the effects we have discussed. Unfortunately, such accurate mass 
measurements might be very difficult to come by for different reasons. 
In the focus point region all scalars will be heavy, leading to small 
production cross section at the LHC. In the co-annihilation line with 
a nearly degenerate stau and a neutralino, the stau decays produce 
very soft taus, which are hard for the LHC to measure. And the higgs 
funnel extends, depending on $\tan\beta$ and $M_T$, to very large values 
of ($m_0,M_{1/2}$), at least partially outside the LHC reach. 
Nevertheless, DM provides in principle an interesting constraint on 
the (supersymmetric) seesaw explanation of neutrino masses, if seesaw 
type-II is realized in nature, a fact which to our knowledge has not 
been discussed before in the literature.

\section*{Acknowledgements}

Work supported by Spanish grants FPA2008-00319/FPA and Accion
Integrada HA-2007-0090 (MEC).  The work of J.N.E.. is supported by {\it
  Funda\c c\~ao para a Ci\^encia e a Tecnologia} under the grant
SFRH/BD/29642/2006. The work of J.C.R. and J.N.E. is also supported by
the RTN Network MRTN-CT-2006-035505 and by {\it
  Funda\c c\~ao para a Ci\^encia e a Tecnologia} through the projects
CFTP-FCT UNIT 777 and   CERN/FP/83503/2008.
W.P.~is partially supported by the German Ministry of Education and 
Research (BMBF) under contract 05HT6WWA and by the DAAD project number 
D/07/13468.

\end{document}